\documentclass[reprint,amsmath,amssymb,aps,prb,nofootinbib]{revtex4-2}
\usepackage{lipsum}
\usepackage{subfiles}

\usepackage{graphicx}
\usepackage{dcolumn}
\usepackage{bm}
\usepackage{booktabs}
\usepackage{textgreek}
\usepackage{appendix}
\usepackage{xcolor}

\newcommand{\sa}{^{[\alpha]}}
\newcommand{\sab}{^{[\alpha,\beta]}}
\renewcommand{\sb}{^{[\beta]}}
\newcommand{\TYPE}{^{[\mathrm{type}]}}
\newcommand{\TYPEij}{^{[\mathrm{type}](i,j)}}

\newcommand{\ijab}{^{(i,j)[\alpha,\beta]}}
\newcommand{\LL}{\mathcal{L}}
\newcommand{\drho}{\varrho}
\newcommand{\bdrho}{\boldsymbol{\varrho}}
\newcommand{\brho}{\boldsymbol{\rho}}
\newcommand{\bv}{\boldsymbol{v}}
\newcommand{\br}{\boldsymbol{r}}
\newcommand{\bdr}{\boldsymbol{\Delta r}}
\newcommand{\brt}{\tilde{\boldsymbol{r}}}
\newcommand{\bdrt}{\boldsymbol{\Delta}\tilde{\boldsymbol{r}}}
\newcommand{\bb}{\boldsymbol{b}}
\newcommand{\bn}{\boldsymbol{n}}
\newcommand{\bt}{\hat{\boldsymbol{t}}}
\newcommand{\bxi}{\boldsymbol{\xi}}
\newcommand{\FPK}{\boldsymbol{F}_\mathrm{PK}^{[\alpha]}}
\newcommand{\Fsub}[1]{\boldsymbol{F}_\mathrm{#1}^{[\alpha]}}
\newcommand{\ens}[1]{\left\langle #1 \right\rangle}
\newcommand{\rot}{\boldsymbol{R}^{[\alpha,\beta]}}

\begin{document}
\title{Dislocation correlations and the continuum dynamics of the weak line bundle   ensemble }
\author{Joseph Pierre Anderson}
\affiliation{School of Materials Engineering, Purdue University, West Lafayette, IN 47907, USA. \\(Currently with Department of Physics and Astronomy, Carthage College, Kenosha, WI 53140, USA.)}
\author{Anter El-Azab}
\affiliation{School of Materials Engineering, Purdue University, West Lafayette, IN 47907, USA.}
\begin{abstract}
Progress toward a first-principles theory of plasticity and work-hardening is currently impeded by an insufficient description of dislocation kinetics, that is, of the dynamic effect of driving forces in a given dislocation theory. Current continuum theories of dislocation kinetics are often incapable of treating the short-range interaction of dislocations. This work presents a kinetic theory of continuum dislocation dynamics in a vector density framework which takes into account the short-range interactions by means of suitably defined correlation functions. The weak line bundle ensemble of dislocations is defined, whereby the treatment of dislocations by a vector density is justified. It is then found by direct averaging of the dislocation transport equation that additional driving forces arise which are dependent on the dislocation correlations. A combination of spatial coarse-graining and statistical averaging of discrete dislocation systems is then used to evaluate the various classes of tensorial dislocation correlations which arise in the line bundle kinetic theory. A novel, chiral classification of slip system interactions in FCC crystals is introduced in order to define proper and improper rotations by which correlation functions corresponding to six interaction classifications can be evaluated. The full set of these six dislocation correlations are evaluated from discrete data. Only the self-correlations (for densities of like slip system) are found to be highly anisotropic. All six classes of correlation functions are found to be of moderate range, decaying within 2-4 times the coarse-graining distance. The correlations corresponding to the coplanar interactions are found to be negligible. Implications of the evaluated correlations for the implementation of vector density continuum dislocation dynamics are discussed, especially in terms of an additional correlation component of the driving force and a gesture toward a coarse-grained dislocation mobility.
\end{abstract}
\maketitle

\section{Introduction}
Since the discovery of dislocations as the mediators of plastic deformation almost a century ago \cite{Orowan1934,Taylor1934,Polanyi1934}, dislocation theory has been an important topic of interest in physics, mechanics, and materials science. Due to the role of dislocations in determining the plastic strength of metals and alloys, a major part of the theory of dislocations was quickly developed, focusing on the mechanics \cite{Burgers1939}, interactions \cite{Peach1950}, and reactions of individual dislocations \cite{Thompson1953} in various crystal structures. However,  the initial question that dislocations were posited to answer remains unsolved even today: precisely \emph{how} do dislocations give rise to work hardening in metals? The development of a first-principles theory of work hardening is still underway because it is a problem of the collective motion of dislocations. Collective motion always involves a certain set of complexities, but dislocation theory comes with its own peculiar difficulties. 

To begin to sort out the types of difficulties faced in treating the collective dynamics of dislocations, let us first distinguish between dislocation \emph{kinematics} and \emph{kinetics} \cite{Hochrainer2007,Zaiser2015}. In treating a collection of dislocations, one often describes the collection by means of a reduced representation. This commonly involves  some set of density measures. Constructing a theory of dislocation \emph{kinematics} involves defining these density measures and their evolution under a prescribed velocity field. There are several such theories which are appropriate for different purposes, as we will discuss below. However, within a given kinematic framework, one will have to define driving forces \cite{Peach1950} and mobilities \cite{kubin2013} whereby the velocity field can be related to the density state. This closure is the problem of dislocation \emph{kinetics}. The fundamental question of both problems is one of scale. Dislocations as line objects, as well as their interactions, are associated with a fine scale near the lattice spacing (on the order of a few Burgers vectors, or ~1 nm). Descriptors of a collection of dislocations are necessarily associated with a coarser scale, be it as small as a few tens of nm or as large as several μm. The problem of dislocation kinematics is associated with finding appropriate dislocation density measures which capture the relevant dislocation information on the desired coarse scale \cite{Bertin2019}. The problem of dislocation kinetics, which is the focus of the present work, is in defining the short-range dislocation interactions—which naturally occur at the fine scale—in terms in terms of the density measures at the coarse scale. After situating our approach to the kinematic problem, we will then proceed to treat dislocation kinetics for this particular kinematic framework.  

Dislocations are topological defects in crystals \cite{Mermin1979}. A volume containing a dislocation will be recognizable by means of the dislocation’s topological invariant, analogous to a ‘charge’ of the dislocation. This topological invariant is the dyad formed by its line tangent and Burgers vector. When this is summed over all dislocations in a given volume, we obtain the Kröner-Nye tensor \cite{Nye1953,dewit1973}. The long-range mechanical fields (and thus dislocation interactions) are produced by the geometric incompatibility which this tensor represents \cite{Bertin2019,dewit1973}. However, the signedness of this measure is the fundamental stumbling block of collective theories of dislocation motion: in summing the topological invariant over a collection of dislocations multiple kinds of cancellation can occur, essentially losing track of some portion of the dislocation population in the volume \cite{Kroner2001}. The portion contributing to the Kröner-Nye tensor is commonly referred to as the geometrically necessary dislocation content (GND). Various kinematic theories define additional density measures in order to account for the cancelled dislocation content—commonly referred to as the statistically stored dislocation content (SSD)—with various levels of precision.\footnote{While in this work we will refer to geometrically necessary or statistically stored dislocations in the plural, we remind the reader that these are net effects of a collection, and a geometrically necessary or statistically stored character is not attributable to a single dislocation.}

    We will discuss two different kinematic theories which treat SSDs in different ways. While there are kinematic theories which model only the evolution of GNDs \cite{Acharya2006,Roy2006}., it is difficult to interpret the velocity associated with this evolution in terms of the velocities of individual dislocations. As a result, kinetic closure of such theories is performed by free-energy or phenomenological arguments rather than coarse-graining arguments \cite{Acharya2006a}. When considering the problem of statistical storage of dislocations, however, a certain distinction immediately arises corresponding to two modes by which the topological invariant might cancel. Firstly, the presence of dislocations of different slip system in the volume leads to cancellation by means of the Burgers vector and the representative plane of possible dislocation tangents \cite{Arsenlis1999}. We will refer to this SSD content as multi-slip SSDs. However, cancellation can also occur due to dislocations on a single slip system by cancellation of the tangent vector \cite{Hochrainer2007}. We will refer to this SSD content as single-slip SSDs. This can arise either due to nontrivial curvature of the dislocations on the scale of the representative volume or due to multiple non-parallel dislocations passing through the volume. { In a curved segment, the component of the length perpendicular to the end-to-end distance is lost. In the extreme case, a full dislocation loop would be entirely stored in this mode.} To create a theory that resolves multi-slip SSDs but not single-slip SSDs, one must consider the density of dislocations on each slip system and their average direction (either in the form of an angular parameter \cite{Sedlacek2010} or in the form of a vector density \cite{Mura1963,Lin2020,Lin2021,Anderson2022}), resulting in two kinematic variables per slip system. We will refer to this as the \emph{vector density framework}. If one wishes to resolve even single-slip SSDs, it is necessary to treat the dislocations as having some distribution over angular space as well as to treat a curvature density \cite{Hochrainer2007}. Alternatively, this treatment may be closed at lower order by considering only the first several coefficients in a series expansion of this distribution \cite{Hochrainer2015}. We will refer to this as the \emph{higher order kinematic framework}. Closure at second order (already requiring approximation \cite{Sandfeld2015,Monavari2016}) requires the treatment of the total line length in the volume, the average curvature of those lines, the average line tangent, and the average of the dyad formed by the line tangent with itself. These are described by one, one, two, and two variables, respectively, for a total of six kinematic variables per slip system when all vector quantities are confined to the glide plane of dislocations. 

Regardless of the kinematic framework chosen, \textit{kinetic} closure, as will be the focus of the present work, amounts to determining relations between the structure at two scales. The relative arrangement of the dislocations at the discrete scale results in all the interaction fields, forces, and, eventually, velocities of interest in the dynamics. Some of this relative arrangement information is preserved in the continuum density field; this is why the coarse-grained Kröner-Nye tensor gives rise to the long-range stress field. This tradeoff between short- and long-range portions of the stress field is at the heart of the fast Fourier methods which have gained recent popularity in the discrete dislocation dynamics community \cite{Brenner2014,Bertin2015,graham2016,Bertin2019,Morin2019}. In such a case, one solves the long-range interactions using the Kröner-Nye tensor while retaining recourse to the discrete dislocation positions to compute the short-range interactions overlooked by the Kröner-Nye tensor \cite{Vivekanandan2022}. In the continuum case, however, this short-range stress contribution must be considered in an average sense. Consideration of the average short-range dislocation structure is performed by analysis of the correlation functions. These allow the relative arrangement information of the coarse-grained density field to be corrected in order to approximate more closely the relative arrangement information in the discrete system. For the higher order theory, where the local dislocation arrangement is more general, this correction is a complex object involving multiple contributions from the connectedness of the lines and even something like a local microstructure \cite{Hochrainer2022}. For the vector density theory, the form is simplified and the correlation largely encodes line connectivity information and the average structure of parallel line bundles \cite{Anderson2021}. 

The kinetic analysis and correlation functions in the present work will be entirely situated in the vector density framework. However, if there is a more complete framework (the higher order framework), why would we bother with the vector density framework? The short answer is parsimony: the vector density framework requires two variables per slip system (24 variables for FCC metals with 12 slip systems), while the higher order framework requires thrice as many variables (72 variables for FCC). This lack of parsimony becomes more of an issue when kinetics begin to be considered. Kinetic closure, as we will describe in more detail shortly, will require a description of average quantities which are quadratic in the density variables (e.g. total elastic energy \cite{Zaiser2015}). These pair densities will have to bridge the two relevant length scales: they relate the relative arrangement information on the order of the discrete dislocations to the relative arrangement stored in the coarse density field. The difference between these relative arrangements, in the form of correlation functions, must be evaluated for every combination of density field variable \cite{Zaiser2015,Hochrainer2022}. A threefold increase in variables becomes a ninefold increase in kinetically relevant variables. Of course, if that is the price for accurately modelling the physical system, so be it. However, what if this increase in variables gained no new information regarding the dislocation configuration? That is, what if you are considering a scale on which physical dislocation arrangements do not contain single-slip SSDs? We know that below some critical distance, considered to be on the order of 50-100 nm \cite{Lin2020,Essmann1979,Xia2015a}, opposite signed dislocations tend to annihilate. At or below this scale, the additional variables become irrelevant, as collections of dislocations on like slip system would bear more resemblance to bundles of parallel lines.\footnote{This assumes the absence of significant dislocation dipole populations, which would represent non-annihilating single-slip SSDs.} We will more rigorously define this weak line bundle regime, where the line bundle assumption holds in a specific sense, in more detail below. 

Work on the dislocation correlation functions has only just begun in recent years. While several works have attempted to outline kinetic frameworks for continuum dislocation dynamics by appealing to dislocation correlations \cite{Zaiser2015,Groma2021,Hochrainer2016}, only just recently have the correlation functions themselves been studied in three dimensions \cite{Anderson2021,Hochrainer2022}. Here we approach the kinetic portion of a vector density theory in a more direct manner, demonstrating how the correlation functions arise naturally by directly averaging the dislocation transport equation. {In so doing, we will find that an additional driving force for dislocation motion arises. This additional non-local force, which we call the correlation force, bears a similarity to the correlation-based effective stresses found in two-dimensional models of dislocation motion \cite{Groma2016,Ispanovity2020}. This previous work in two-dimensions, however, only considers a local density approximation to this force \cite{Zaiser2015}, considered here in a more general non-local form.}
Also, whereas the earlier investigations of correlation functions only treat a small subset of the relevant correlations, here we will evaluate all correlations which can in principle arise in the vector density framework.

In the remainder of this work, we present a kinetic theory of dislocation motion in the vector density framework. In section II, we outline the transport relationship for discrete dislocations and how this is coarse-grained in the line bundle regime. In this way, we show how correlation functions naturally arise and close the kinetic equations. In section III, we present a means by which these correlation functions might be evaluated from discrete dislocation data. This is followed in section IV by a symmetry analysis of slip systems in FCC crystals which reduces the number of unique correlation functions which must be considered between pairs of slip systems. In sections V and VI we present the calculations and their results, followed in section VII by a discussion of the implications of these findings for continuum theories of dislocation motion.

\section{Discrete and Coarse-grained Kinetics of Dislocation Transport}
Beginning with a treatment of discrete dislocation dynamics and the coarse-graining process, we will show how dislocation correlation functions arise in the kinetics of coarse-grained dislocation dynamics. Upon application of the (weak) line bundle assumption, we show how the correlation functions enter into the dynamics in a straightforward manner. 

\subsection{Discrete Dislocation Dynamics}
Let us begin with a treatment of the underlying discrete dislocation lines before introducing the coarse-graining treatment. The dislocation system in an FCC crystal can be described as a collection of twelve line objects (1-manifolds) $\LL\sa$  embedded in a crystal space. $\mathcal{M}$ which is roughly equivalent to $\mathbb{R}^3$.\footnote{Although deformation \textit{per se} contradicts the equivalence between the crystal manifold $\mathcal{M}$ and $\mathbb{R}^3$, we note that for short-range effects (as will be examined in this work), the crystal manifold is approximately flat.}  These embedded line objects also define two measures on the crystal manifold, namely a scalar line length $\Lambda\sa$  and the geometric dislocation content $\boldsymbol{\kappa}\sa$  in a given region $\Omega \subseteq \mathcal{M}$:
\begin{align}
    \Lambda^{[\alpha]}(\Omega \subseteq \mathcal{M})&:=\int_{\LL\sa\cap \Omega} dl,\\
    &:= \int_\Omega \drho\sa (\br)d^3\br,\\
    \boldsymbol{\kappa}\sa(\Omega \subseteq \mathcal{M})&:=\int_{\LL\sa\cap \Omega} \bxi^{[\alpha]}(\br_l)dl,\\
    &:=\int_\Omega \bdrho\sa(\br)d^3\br,
\end{align}
where we have denoted by $\bxi^{[\alpha]}(\br)$ the unit direction of the differential line element $dl$. While the scalar line content $\Lambda\sa$ will still be of some use in the analysis of discrete dislocation data, the theoretical formalism is concerned much more with the geometrically necessary dislocation content. These measures imply the existence of a singular distribution or density $\bdrho\sa$. This singular density is analogous to a Dirac distribution concentrated on the line object $\LL\sa$. This distribution allows an identification of the discrete line object with a singular density field, or, in the language of differentiable manifolds, with a two-form. The nature of this density as a two-form allows us to leverage a result from the theory of differentiable manifolds, namely that the evolution of the lines is given by a Lie derivative with respect to some velocity field $\bv\sa$:
\begin{align}
    \dot{\bdrho}\sa &= \nabla \times \left(\bv\sa\times \bdrho\sa\right)-\bv\sa\left(\nabla \cdot \bdrho\sa\right),\label{eq:trans_openline}\\
    &=\nabla \times \left(\bv\sa\times \bdrho\sa\right).\label{eq:trans_closedline}
\end{align}
We have chosen not to burden the reader with the interior product and exterior derivative notation. Rather we will opt for the above vector calculus notation, albeit with the caveat that the particular action of the gradient operators on the singular density is to be understood as an exterior derivative.\footnote{As a matter of fact, dislocations are not most perfectly represented by singular densities. Rather, they are better represented by well-defined continuum densities resolved on the order of the lattice spacing \cite{Cai2006,Po2018}. In such a case, the gradient operators used in Eqs. (\ref{eq:trans_openline},\ref{eq:trans_closedline}) require no qualification.}  The transport equation for open lines [Eq. (\ref{eq:trans_openline})] has three terms: the first describes transport due to dislocation glide, while the second describes the motion of the endpoints of the lines. In the case where the dislocation line object $\LL\sa$ is a collection of closed curves, the divergence (exterior derivative) of $\bdrho\sa$ is null, corresponding to the physical fact that dislocation lines cannot terminate in the crystal. For a treatment of systems where assumption is relaxed to treat the closure failure created in connected dislocation networks (e.g.  by way of cross-slip, junction reactions, etc.), see \cite{Starkey2022}. In the present work we wish only to treat the kinetics of dislocation glide, and so ignore the endpoint transport, resulting in Eq. (\ref{eq:trans_closedline}).

The above treatment is a kinematic description of the evolution of discrete line objects in space. In order to describe a kinetic theory of dislocation motion, we must include a physically motivated prescription of the velocity field $\bv\sa$. In discrete models of dislocation dynamics, the motion of the dislocation lines is assumed to be overdamped and therefore linearly related to the stress field evaluated on the dislocation line \cite{kubin2013,Cai2004,Sills2016a}. The form of the velocity law is given by the following simplified functional notation:
\begin{equation}
    B\bv\sa(\br):= \FPK\left[\bxi\sa(\br)\otimes\bdrho\sb(\br')\right].
\end{equation}
Here $B$ is a material specific drag parameter, and the Peach-Koehler functional $\FPK$ is a linear, vector-valued functional. As a result, the brackets should be understood to denote an integration over $\br'$. From the details of how this Peach-Koehler functional is derived, cf. appendix 1. Its form is given by:
\begin{widetext}
    \begin{equation}
        \left(\FPK\left[\bxi\sa(\br)\otimes\bdrho\sb(\br')\right]\right)_l := \epsilon_{jkl} b_i\sa \xi_k\sa (\br) \left( \sigma^0_{ij} +\sum_{\beta=1}^{N_S} \int_{\mathcal{M}} \hat{\sigma}_{ijm}\sb (\br-\br') \drho_m\sb(\br') d^3\br' \right), \label{eq:PK_functional}
    \end{equation}
\end{widetext}
where $b\sa_i$ is the Burgers vector of slip system $[\alpha]$, $\sigma^0_{ij}$ the externally applied stress,  $\hat{\sigma}_{ijm}\sa(\bdr)dl_m$ is the stress field surrounding a differential dislocation segment $dl_m$,{ and $\epsilon_{jkl} $ is the standard Levi-Civita tensor}. The summation runs over all $N_S$ slip systems considered in the crystal.

As simple as they may seem, Eqs. (\ref{eq:trans_closedline}-\ref{eq:PK_functional}) amount to a complete theory of the glide of discrete dislocations. Discrete frameworks consist primarily in evaluating interaction forces [Eq. (\ref{eq:PK_functional})], and updating segment positions, which due to the line expansion inherent in Eq. (\ref{eq:trans_closedline}) is not a trivial operation \cite{Sills2016a}. We now turn our attention for the remainder of this section to the evolution of continuum densities of dislocations. To do so, we will outline a class of averaging operators which allow the discrete dislocation kinetics to be coarse-grained rather straightforwardly. We will then show how correlations naturally enter into such a kinetic theory.

\subsection{Continuum Dislocation Dynamics \\in Line Bundle Ensembles}
Before showing how the discrete transport kinetics are averaged, we first must define what we mean by an averaging operator. We will denote this average with angle brackets $\ens{\cdot}$. Its main quality of interest is that it takes discrete densities to a smooth continuum density field:
\begin{equation}
    \brho\sa(\br) := \ens{\bdrho\sa(\br)}.
\end{equation}
Additionally, the average is a linear projection operator. That is, it has the following properties for constants $c,d$ and linear functionals of the line configuration $A,B$:
\begin{align}
    \ens{cA(\LL) + d B(\LL)} &= c\ens{A} + d \ens{B}, \\
    \ens{\ens{A}} &= \ens{A}.
\end{align}

Using only these properties, it is possible to derive a kinematic theory of dislocation motion in a continuum sense \cite{Hochrainer2007,Hochrainer2015}. However, deriving a kinetic theory becomes difficult without making use of further assumptions \cite{Zaiser2015}. These are chosen by considering what type of dislocation states are considered in the average, analogously to the operation in statistical mechanics of selecting an ensemble. In the present case, we choose to utilize the so-called line-bundle assumption \cite{Anderson2021,Anderson2022}. This corresponds to an assumption that the dislocation system is highly polarized on the representative distances on which the density field is modelled. That is to say, this assumption corresponds to the case of few single slip SSDs. For clarity, we will define two variations of this assumption: a strong line bundle assumption and a weak line bundle assumption.

In our previous work with correlations in line bundle ensembles \cite{Anderson2021}, we utilized a strong form of the line bundle assumption, which states that for any constant vector $\boldsymbol{a}$,
\begin{equation}
    \ens{\lvert \boldsymbol{a} \cdot \bdrho\sa(\br)\rvert}= \lvert\boldsymbol{a}\cdot \brho\sa(\br)\rvert.
\end{equation}
An equivalent statement would be to say that fluctuations in the line direction,
\begin{equation}
    \delta \bxi\sa(\br):=\bxi\sa(\br)-\hat{\brho}\sa(\br)
\end{equation}
--where $\hat{\brho}\sa$ represents the unit direction of $\brho\sa$--are by definition null. This strong line bundle assumption has the benefit that it produces convenient properties as corollaries of this assumption, especially in the form of the correlation functions. However, this is a rather stringent condition which in practice is rarely fulfilled in physical dislocation systems.

Nonetheless, appropriate averaging processes can be defined where the line bundle assumption holds approximately. As a result, we introduce a weak line bundle assumption. Rather than defining fluctuations in line direction as null, we assume they are simply small to second order: $\delta  \xi_i^2 \approx 0$. In this case, any quantities second order or higher in the line direction can factor out the average line direction once:\footnote{Technically in the weak line bundle approximation [Eq. (\ref{eq:WLBA})], the right-hand-side should be fully symmetrized. This is omit ted for brevity. }
\begin{equation}\label{eq:WLBA}
    \ens{\bxi^{\otimes n} A(\LL)} \approx \hat{\brho}\ens{\bxi^{\otimes (n-1)} A(\LL)}.
\end{equation}

Whereas the result of the strong line bundle assumption is that the dislocation system is perfectly polarized ($|\brho|=\ens{\drho}$), the weak line bundle assumption simply operates in the high polarization regime $|\brho|/\ens{\drho}\approx 1$. In the following section, we will examine a particular averaging procedure which satisfies this property, but at this time we will simply examine the effects of a weak line bundle averaging operator on the kinetics of discrete dislocation transport.

Because the averaging operator is linear, it commutes with derivatives and other linear operators. As a result, averaging both sides of Eq. (\ref{eq:trans_closedline}) results in the following coarse-grained transport equation: 
\begin{equation}
    \dot{\brho}\sa = \nabla \times \ens{\bv\sa \times \bdrho\sa}. \label{eq:trans_cont}
\end{equation}
As is evident from this equation, a solution of the dislocation dynamics amounts to finding a closed-form expression for the cross product of the velocity and density vector fields. We do this by introducing the relation for the velocity field to obtain:
\begin{align}
    B\ens{\bv\sa \times \bdrho\sa} &= \ens{\FPK\left[\bxi\sa(\br)\otimes \bdrho\sb(\br')\right]\times \bdrho\sa(\br)} \nonumber \\
    &= \ens{\FPK\left[\bdrho\sa(\br)\otimes \bdrho\sb(\br')\right]\times \bxi\sa(\br)} \nonumber \\
    &\approx \ens{\FPK\left[\bdrho\sa(\br)\otimes \bdrho\sb(\br')\right]} \times \hat{\brho}\sa(\br) \nonumber\\
    &= \FPK\left[\ens{\bdrho\sa(\br)\otimes \bdrho\sb(\br')}\right] \times \hat{\brho}\sa(\br) \label{eq:trans_2ptdens}
\end{align}
by the application of, respectively, the definition of the unit direction, the weak line bundle assumption, and the linearity of the average and Peach-Koehler functional.

In the final relation above, a new quantity of interest has arisen: the two-point density { $\rho^{[\alpha,\beta]}(\boldsymbol{r},\boldsymbol{r'})$ which is given by},
\begin{equation}\label{eq:17}
    \brho\sab(\br,\br') :=\ens{\bdrho\sa(\br)\otimes \bdrho\sb(\br')}.
\end{equation}
As is common in the analysis of density hierarchies, we consider this average to be expressed without loss of generality by the product of the single-point densities and a correlation function $d\ijab(\br,\br')$. This correlation function expresses the scalar correlation between each respective tensor component of the product density for each combination of slip systems. That is, 
\begin{equation} \label{eq:corr_def}
    \rho\sab_{ij}(\br,\br') :=\rho\sa_i(\br)\rho\sb_j(\br') \left(1 + d\ijab(\br,\br')\right).
\end{equation}
The scalar operation of the correlation function is one way of denoting the diagonal nature of an in general tensorial correlation:
\begin{equation}
    d\sab_{kijl} (\br,\br') := d\ijab(\br,\br')\delta_{ik}\delta_{jl},
\end{equation}
where $\delta_{ij}$ here represents the Kronecker delta. This diagonality property was shown to be a corollary of the strong line bundle assumption \cite{Anderson2021}; we presume that it holds, at least approximately, in the case of the weak line-bundle approximation as well.

Substituting this definition into the transport relation [Eqs. (\ref{eq:trans_cont},\ref{eq:trans_2ptdens})], we obtain:
\begin{align}
    \dot{\brho}\sa &= B^{-1} \nabla \times \left\{  \left( \Fsub{MF} + \Fsub{C}\right) \times \brho\sa \right\}, \label{eq:Kinetic_1} \\
    \Fsub{MF}(\br) &= \FPK\left[\hat{\brho}\sa(\br)\otimes \brho\sb(\br')\right],\\
    \Fsub{C}(\br) &= \FPK\left[\hat{\brho}\sa(\br)\cdot \boldsymbol{d}\sab(\br,\br')\cdot \brho\sb(\br')\right].\label{eq:Kinetic_2}
\end{align}
The mean-field force term, $\Fsub{MF}$, captures the driving force arising due to the eigenstress field generated by the continuum densities $\brho\sb$. Assuming the dislocation correlation functions to be non-trivial, however, we can see that the mean-field solution of the driving force actually misrepresents the dislocation transport. The correlation-dependent force term, $\Fsub{C}$, then serves to correct the dynamics, reincorporating into the theory the short-range dislocation interactions lost in the coarse-graining process. 

We have already arrived at an important result: a kinetic theory for the evolution of the dislocation density vector in highly polarized dislocation systems. If the correlation function can be evaluated from discrete data, the kinetic theory would be complete. In the remainder of this work, we present a method for averaging discrete dislocation data in a way consistent with the line bundle assumption so that the correlation functions can be evaluated from discrete dislocation data.

\section{Averaging of Discrete Dislocation Systems}
We have shown that for certain averaging operators—i.e. those for which the weak line bundle assumption [Eq. (\ref{eq:WLBA})] holds—the correlation functions enter straightforwardly into the evaluation of the dislocation velocity field. In the section that follows, we present a specific averaging operator which fulfills the line bundle assumption and outline an empirical method for evaluating the correlation function from discrete dislocation dynamics. This averaging process consists of a spatial convolution of the dislocation line with a weight function of finite range as well as a statistical homogeneity argument that treats all points in the crystal as equivalent.

\subsection{Spatial coarse-graining of discrete dislocations}
For some discrete dislocation line configuration $\LL_0 $ with corresponding discrete density field $\bdrho_0$, define the continuum density $\brho(\br;L)$ as follows:
\begin{equation}\label{eq:CGrho}
    \brho(\br;L) := \int_{\Omega_L} w_L (\br-\br')\bdrho_0(\br') d^3\br'.
\end{equation}
where $w_L$ is some weight function with compact support $\Omega_L$ characterized by some characteristic coarse-graining length $L$. We note that for this spatially averaged density field, there will obviously be a qualitative difference between the density field as this length tends to 0 and the density field as this length tends to infinity. This qualitative difference is closely related to the emergence of what we have discussed in the introduction as single-slip SSDs. For small coarse-graining lengths, we note that there is actually a small length $L^*$ which is used in discrete dislocation dynamics to remedy the non-physical singularity in the elastic field of dislocations ([38], [39]). That is to say, physical, discrete dislocations are better modelled as density fields of the type $\brho(\br;L^*)$, where $L^*$ is on the order of 5-10 times the Burgers vector magnitude, or about 2-3 nm in copper. Here single-slip SSDs are absent by definition. On the other extreme, it is obvious that at some sufficiently great $L$, at least on the order of 1 {\textmu}m, the volume $\Omega_L$ surrounding any point could contain dislocation loops, or at least dislocations sufficiently curved so as to make the vector cancellation in the spatial average non-negligible, resulting in significant populations of single-slip SSDs. However, we expect to find intermediate lengths, on the order of 20-100 nm, where there is a \textit{de facto} absence of single-slip SSDs. This corresponds to the physical requirement that these single-slip SSDs (opposite signed dislocations in $\Omega_L$) actually annihilate within some critical distance. This critical distance is often estimated to be on the order of ~50 nm \cite{Xia2015}.

A consequence of the above considerations is that the definition of the correlation function in Eq. (\ref{eq:corr_def}) now supports a more precise interpretation. Namely, the correlation encodes some relationship between density products at different scales $L^*$ and $L$. It attempts to recover the fine product densities by means of an appropriate linear relationship with the coarser product densities.

The result of applying this density field calculation to a discrete dislocation configuration is that for some collection of points $\{\br_i\}_{i=1}^{N_r}$, we now have in hand the density field at those points $\brho\sa(\br_i;L)$ for various convolution lengths $L$. However, in terms of measuring the product density, this only amounts to a single measurement of the product density. This is because the product density
\begin{equation}
    \brho\sab(\br_i,\br_j;L) =\brho\sa(\br_i;L)\otimes\brho\sb(\br_j;L).
\end{equation}
has no sense of equivalence between some classes of point pairs. In order to get some sense of the representative behavior of classes of these point pairs, we must define a further averaging process.

\subsection{Ensemble averaging of dislocation product densities}
If we hope to build some picture of the average relationships between product densities at different scales we will need to reduce the space of dependencies of the correlation function. In its form in Eq. (\ref{eq:corr_def}), it depends on a six-dimensional space, a tuple of two positions $(\br,\br')$. However, it is evident that in bulk crystals there is some degree of arbitrariness to any specific choice of origin. If the only local information that the correlation can be dependent on is the dislocation density, then it follows from such a statistical homogeneity argument that the correlation should only be dependent on the separation distance $\bdr:=\br'-\br$ \cite{Anderson2021}. Additionally, we should expect the long range behavior of the dislocation state $\bdr\to\infty$ to be uncorrelated. That is, for some function of the system state at two points $f(\br,\br'):=f_1(\br) f_2(\br')$, we expect
\begin{equation}\label{eq:uncorrelated}
    \lim_{\br'-\br \to \infty}\ens{f(\br,\br')} = \ens{f_1(\br)}\ens{f_2(\br')}.
\end{equation}

In any case, this statistical homogeneity argument raises further questions. Namely, which points do we treat as equivalent and how can we use this equivalence to reason to a correlation function? As should be evident from our kinetic derivation, we are interested in points where there is a dislocation at both $\br$ and $\br'$. Otherwise, there is no transport nor interaction to be spoken of. As a result, we are interested in the atmosphere of discrete density surrounding a point with non-zero discrete density. Considering a sample set $S\sa$ of points where the discrete density is nonzero,
\begin{equation}
    S\sa := \left\{   \br\in\mathcal{M} \; : \; \rho\sa(\br,L^*) > 0    \right\},
\end{equation}
we may then partition the direct product of the sample space into sets of like separation vector in order to find pairs of interest to us:
\begin{equation}
    \Pi\sab(\bdr) := \left\{ (\br,\br')\in S\sa \times S\sb \; : \; \br'-\br = \bdr           \right\}. \label{eq:pairs}
\end{equation}

We will then simply examine the density product data at two scales, $L^*$ and $L$, for position pairs in $\Pi\sab(\bdr)$. That is, we will examine the scattered product density data,
\begin{align}
    \Big\{ \Big(\rho\sab_{ij}(\br,\br';L^*),\; &\rho\sab_{ij}(\br,\br';L)\Big) \nonumber \\ 
    &\forall \;(\br,\br') \in \Pi\sab(\bdr)\Big\},
\end{align}
and fit various correlation functions relating the two.

\section{Interaction Symmetries in FCC Crystals}

While it certainly could be of some utility to fit correlation functions for all unique pairs of slip systems [$\alpha,\beta$], it would not be particularly parsimonious. If all 12 slip systems are considered, there are a total of 144 slip system interactions (78 if these are considered symmetric under interchange of $\alpha,\beta$). Add to that the 4 tensorial components of the correlation for each and things quickly get out of hand. However, it is expected that some of these correlation functions are measuring qualitatively similar behavior. More specifically, it is expected that the spatial features of the correlation functions will be somehow relative to a special coordinate system dependent on the combination of slip systems being considered. As a result, we may wish to revise the collection of equivalent position pairs [Eq. (\ref{eq:pairs})] to also treat certain classes of slip system interactions as equivalent. 

With this motivation in mind, we take a brief digression from the main investigation of correlation functions to examine relationships between slip systems, and from those relationships to form certain equivalence classes. These equivalence classes will allow us to define distinct classes of correlation functions which can be averaged together in the sense outlined in the previous section. For the present work, we consider only the case of FCC crystals, for which slip systems are commonly defined in terms of the faces and edges of the so-called Thompson tetrahedron \cite{Thompson1953}. In this section, we will examine the algebraic structure of the Thompson tetrahedron and their resulting equivalence classes. Some of the results will be standard--the relationships of various classes of slip systems are well known in the theory of dislocation reactions in FCC crystals \cite{Lin2020,Hull2011}--but we will also make a distinction which to our knowledge is novel, namely that certain interactions have an associated handedness or chirality. In doing so we reduce the number of slip system interactions which must be considered from 144 (or 78) to 6. 

This classification problem to which this section is devoted can be defined as follows. Given two slip systems $\alpha,\beta$, let us find:
\begin{enumerate}
\item all other systems $\alpha^\prime,\beta^ \prime $ such that the resulting dislocation reaction is of the same type,
\item a (possibly improper) coordinate transformation such that the dihedral planes and Burgers directors of the two slip systems are mapped onto constant positions in space, and 
\item signs according to which the Burgers vectors—and accordingly, the dislocation line sense—should be flipped in order to map the Burgers vectors onto constant positions.
\end{enumerate}

In order to devise such a scheme, we will first outline the algebraic structure of the 12 slip systems in FCC crystals, augmenting the well-known algebra of the twelve perfect dislocations in FCC crystals described by the Thompson tetrahedron \cite{Thompson1953}. This algebra will then allow us to enumerate six general classifications of slip system interactions and the transformations to which they naturally give rise. These classifications correspond to the traditional classification according to the reaction process which takes place between them, i.e. self, cross-slip, coplanar, glissile junction, Lomer-type, and Hirth-type interactions. We will also find that the latter four classes can be further divided into  left- and right-handed subclasses. 

\subsection{Algebraic operations on the Thompson tetrahedron}
\subsubsection{Vectors and directors on the Thompson tetrahedron}
To begin, we must first define two sets of quantities, namely a set of slip plane normal vectors $\bn_i$ and Burgers vectors $\bb_i$. These are the faces and edges, respectively, of a tetrahedron. As a result there are four \{111\}-type normal vectors and six ⟨110⟩-type Burgers vectors, listed in Table \ref{tab:BNdef}\footnote{{Note that these vectors are all defined rationally, and thus do not have unit magnitude.}}. The senses of the normals can be well-defined as the outward-facing normal vectors of the tetrahedral faces. However, it is not possible to adequately define a sense of the Burgers vectors, and as a result the sense of these in Table \ref{tab:BNdef} is arbitrarily chosen following \cite{Devincre2011}.

\begin{table}[]
\caption{Normal vectors and Burgers vectors defined.\label{tab:BNdef}}
{%
\begin{tabular}{@{}cc|cc@{}}
\toprule
\toprule
\multicolumn{2}{c|}{Normal vectors}   & \multicolumn{2}{c}{Burgers vectors} \\
Identifier & Plane                    & Identifier      & Direction         \\ \midrule
\textalpha   & $(\bar 1 \bar 1 \bar 1)$ & a               & $[\bar 1 01]$     \\
\textbeta    & $( 1 \bar 1  1)$         & b               & $[011]$           \\
\textgamma   & $( 1  1 \bar 1)$         & c               & $[\bar 1 10]$     \\
\textdelta   & $(\bar 1  1 1)$          & d               & $[110]$           \\
           &                          & e               & $[01\bar 1]$      \\
           &                          & f               & $[1 0 1]$           \\ 
           \bottomrule \bottomrule
\end{tabular}%
}
\end{table}

One can see that the sign of the Burgers vector is irrelevant by considering that the topological invariant of a dislocation with line sense $\boldsymbol{\xi}$ and Burgers vector $\bb$ is the outer product $\boldsymbol{\xi}\otimes \bb$. Therefore the dislocation is preserved upon reversal of the sign of the Burgers vector so long as the line sense is also reversed. Thus it is more useful to our present purposes to consider Burgers directors, which are non-polar vectors commonly used in the theory of liquid crystals \cite{Allen1999}. To consider vectors as directors, we need an operation which can eliminate the sign of the Burgers vectors. We will denote this sign product as a dot product $b_i\cdot b_j$:
\begin{equation}
b_i\cdot b_j := \frac{\bb_i\cdot \bb_j}{|\bb_i\cdot \bb_j|} \in \{\pm 1\}
\end{equation}
This operation returns the ‘alignedness’ of one Burgers vector with respect to another, and as such is commutative. The multiplication table of this sign product operator is shown in table II. This operation also has the expected property that the resulting sign is sensitive to the sign reversal of either input vector:
\begin{equation}
b_i\cdot (-b_j) =(-b_i)\cdot b_j = -(b_i\cdot b_j)
\end{equation}
and so is well-defined for the flipped Burgers vectors as well.
\begin{table}[h]
\caption{Dot product of Burgers vectors\label{tab:Dot}}
{
\begin{tabular}{@{}c|cccccc@{}}
\toprule \toprule
  & a & b & c & d & e & f \\ \colrule
a & + & + & + & - & - & 0 \\
b & + & + & + & + & 0 & + \\
c & + & + & + & 0 & + & - \\
d & - & + & 0 & + & + & + \\
e & - & 0 & + & + & + & - \\
f & 0 & + & - & + & - & + \\ \bottomrule \bottomrule
\end{tabular}%
}

\end{table}

\subsubsection{Wedge products of normal vectors and Burger directors}
The two algebraic operations which will define the equivalence classes of slip systems are as follows. There are natural operations which arise in the discussion of planes and vectors. First, an operation that takes two normal vectors and returns the dihedral direction: the oriented line direction along which the planes intersect. Secondly, an operation that takes two Burgers directors and returns the plane to which they both belong. We will define both of these as a wedge product. The wedge product of two normal vectors $n_i\wedge n_j$ is defined as 
\begin{equation}
n_i\wedge n_j := \frac{1}{2}(\bn_i\times\bn_j) \in \{\pm b_k\}
\end{equation}
and returns a direction which is a valid Burgers vector. The multiplication table for this operation is shown in Table \ref{tab:normalwedge}; one can see that it inherits anticommutativity from the cross product. While this product will be of use for defining a preferred polarity of the Burgers vectors, it does not lead to interesting classifications of slip system interactions. 

\begin{table}[h]
\caption{Normal vector wedge product\label{tab:normalwedge}}
{
\begin{tabular}{c|cccc}
\toprule
\toprule
& \textalpha & \textbeta & \textgamma & \textdelta \\ \colrule
\textalpha & 0        & a       & -c       & e       \\
\textbeta  & -a       & 0       & b        & -d      \\
\textgamma & c        & -b      & 0        & f       \\
\textdelta & -e       & d       & -f       & 0        \\ 
\bottomrule
\bottomrule
\end{tabular}%
}
\end{table}

Additionally, the wedge product $b_i \wedge b_j$ is defined as
\begin{equation}
b_i\wedge b_j :=(b_i \cdot b_j ) \bb_i \times \bb_j \in \{\pm n_k\}
\end{equation}
and returns a directed slip plane normal vector. The multiplication table for this operation is shown in table IV. The result of this product gives the slip plane to which the aligned Burgers vectors belong in a right-handed sense. Consistently with this interpretation, we see that this product is also anti-commutative, but it is insensitive to the sign of either Burgers vector:
\begin{align}
b_i \wedge b_j &= -(b_j \wedge b_i ) \\
&= (-b_i) \wedge b_j \\
&= b_i \wedge (-b_j) \\
&= (-b_i) \wedge (-b_j) 
\end{align}
The Burgers wedge product gives two interacting slip systems a well-defined chirality with respect to each other. As we will see below, this chirality actually represents an important distinction within the traditional FCC slip system interaction classes, and will be essential to defining equivalent coordinate systems as well.

\begin{table}[h]
\caption{Burgers director wedge product\label{tab:Burgwedge}}
{%
\begin{tabular}{c|cccccc}\toprule \toprule
  & a         & b         & c         & d         & e         & f         \\ \colrule
a & 0         & -\textbeta  & \textalpha  & \textbeta   & -\textalpha & 0         \\
b & \textbeta   & 0         & -\textdelta & -\textbeta  & 0         & \textgamma  \\
c & -\textalpha & \textgamma  & 0         & 0         & \textalpha  & -\textgamma \\
d & -\textbeta  & \textbeta   & 0         & 0         & \textdelta  & -\textdelta \\
e & \textalpha  & 0         & -\textalpha & -\textdelta & 0         & \textdelta  \\
f & 0         & -\textgamma & \textgamma  & \textdelta  & -\textdelta & 0        \\ \bottomrule \bottomrule
\end{tabular}%
}
\end{table}
\subsection{Classification of slip system interactions}
Having defined these algebraic operations between slip planes and Burgers directors, we may now define our twelve slip systems. Each allowable combination of slip plane and Burgers vector is considered (i.e. those for which the Burgers and normal vectors are orthogonal), and these arbitrarily numbered combinations are shown in Table \ref{tab:slips}.

\begin{table}[h]
\caption{Slip system definitions. \label{tab:slips}}
{%
\begin{ruledtabular}
\begin{tabular}{@{}c|cccccccccccc@{}}

Slip system    & 1        & 2       & 3       & 4        & 5        & 6        & 7        & 8       & 9        & 10       & 11       & 12       \\ \colrule
Burgers vector & a        & a       & b       & b        & c        & c        & d        & d       & e        & e        & f        & f        \\
Normal vector  & \textalpha & \textbeta & \textbeta & \textgamma & \textgamma & \textalpha & \textdelta & \textbeta & \textalpha & \textdelta & \textgamma & \textdelta \\
\end{tabular}%
\end{ruledtabular}
}
\end{table}

\begin{figure}
\includegraphics[width = 0.8\columnwidth]{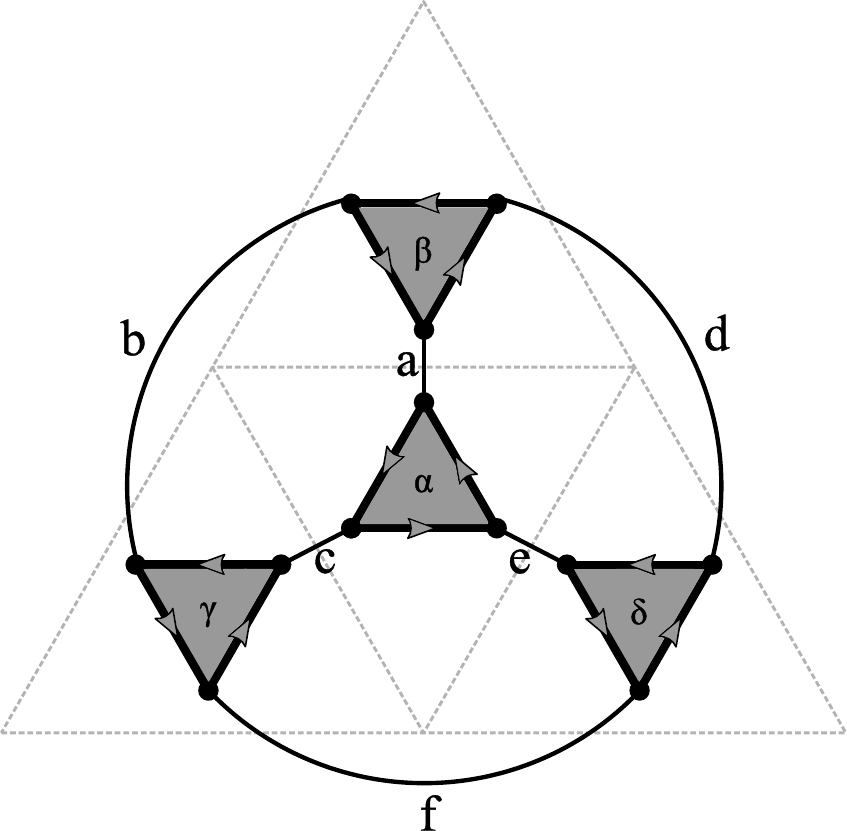}
\caption{The slip system graph structure. Nodes represent slip systems; nodes connected by an undirected edge share a Burgers vector, while nodes connected by a directed edge share a slip plane normal vector. The direction of the edge between two nodes $1\to2$ is chosen such that $b^{[1]} \wedge b^{[2]} =n^{[1]} =n^{[2]}$ . The Thompson tetrahedron is shown faintly superimposed on the graph for reference.\label{fig:key}}
\end{figure}

These slip systems form nodes of a graph structure shown in Fig. \ref{fig:key}. We note that some of the edges of this graph are directed, and some undirected. Two slip systems are connected by a thin, undirected line if they share a Burgers vector. Two edges are connected by a thick directed line if they share a normal vector. The direction of these edges is specified such that for two slip systems connected $1\to 2$, the wedge product of their Burgers directors is right-handed on their shared normal plane; i.e.
\begin{equation}
b^{[1]} \wedge b^{[2]} = n^{[1]}=n^{[2]}
\end{equation}
This graph structure is also shown superposed on the traditional net representation of the Thompson tetrahedron to demonstrate the correspondence.

\subsubsection{Classification conditions}
The classification scheme is as follows. Self-interactions are those for which $\alpha=\beta$, or
\begin{equation}
    \mathrm{self} := \Big\{ [\alpha,\beta] \;:\; n^{[\alpha]} = n^{[\beta]} \; \mathrm{and}\; b^{[\alpha]} = b^{[\beta]}  \Big\}.
\end{equation}

Cross-slip interactions are those slip systems which share a Burgers vector but glide on different slip planes, or:
    \begin{equation}
    \mathrm{xslip} := \Big\{ [\alpha,\beta] \;:\; n^{[\alpha]} \neq n^{[\beta]} \; \mathrm{and}\; b^{[\alpha]} = b^{[\beta]}  \Big\}.
\end{equation}

Coplanar reactions occur when two dislocations on the same slip plane react to form a dislocation segment with a new Burgers vector which can still glide on the original slip plane. This interaction class is given by:
\begin{align}
    \mathrm{coplanar}(\chi) := \Big\{ [\alpha,\beta] \;:\; & b^{[\alpha]} \wedge b^{[\beta]} =\chi n^{[\alpha]} \nonumber \\
    \mathrm{and}\; &n^{[\alpha]} = n^{[\beta]}  \Big\}.
\end{align}
where we have introduced a chirality $\chi=-1,+1$ for left- and right-handed interactions, respectively. 

Glissile junction reactions occur when a similar process occurs between dislocations on different slip systems, resulting in a third glissile segment on one of the two original slip planes. We consider only the interactions where the resulting segment would be glissile on the plane of the second reacting segment.\footnote{As a result of this asymmetric constraint, the glissile junction class is the only class which is not symmetric under interchange of the two slip systems considered.} This corresponds to the case in which the first Burgers vector is along the dihedral direction, i.e. 
\begin{align}
    \mathrm{glissile}(\chi) := \Big\{ [\alpha,\beta] \;:\; &(b^{[\alpha]} \wedge b^{[\beta]})  =\chi n^{[\beta]} \nonumber\\ 
    \mathrm{and}\; &  n\sa \neq n\sb  \Big\}.
\end{align}

Lomer locks are a sessile junction segment formed when two dislocations react to form a valid ⟨110⟩-type Burgers vector, but there does not exist a tetrahedral face which contains both the dihedral direction (the line direction of the resulting segment) and the resulting Burgers vector \cite{Lomer1951,Cottrell2010}. For a right-handed Lomer-type interaction, the dihedral direction $n^{[\alpha]} \wedge n^{[\beta]}$   is right-handed w.r.t. $b^{[\alpha]}$  and left-handed with $b^{[\beta]}$, i.e. 
\begin{align}
    \mathrm{lomer}(\chi) := \Big\{ [\alpha,\beta] \;:\; &(n^{[\alpha]} \wedge n^{[\beta]}) \wedge b^{[\alpha]} =\chi n^{[\alpha]} \nonumber\\ 
    \mathrm{and}\; &b^{[\beta]} \wedge (n^{[\alpha]} \wedge n^{[\beta]}) =\chi n^{[\alpha]}  \Big\}.
\end{align}

Lastly, Hirth locks are a sessile junction formed when dislocations with orthogonal Burgers vectors react, forming a dislocation segment with a $\langle$100$\rangle$-type Burgers vector \cite{Hirth1961}. These segments are sessile because this is not a close-packed direction and thus the energy barrier for slip along this direction is prohibitively high \cite{Hull2011}. Although it is possible to use the requirement $b_i\cdot b_j=0$ to identify Hirth-type interactions, this definition would ignore the chiral nature of the interaction. Rather, we define it similarly to the Lomer-type interactions, with the distinction that for a right-handed Hirth-type interaction, the dihedral direction $n^{[\alpha]} \wedge n^{[\beta]}$   is right-handed w.r.t. both $b^{[\alpha]}$  and $b^{[\beta]}$:
\begin{align}
    \mathrm{hirth}(\chi) := \Big\{ [\alpha,\beta] \;:\; &(n^{[\alpha]} \wedge n^{[\beta]}) \wedge b^{[\alpha]} =\chi n^{[\alpha]} \nonumber\\
   \mathrm{and}\;  &(n^{[\alpha]} \wedge n^{[\beta]}) \wedge b^{[\beta]} \wedge=\chi n^{[\alpha]}  \Big\}.
\end{align}

\begin{figure}
\includegraphics[width =86mm]{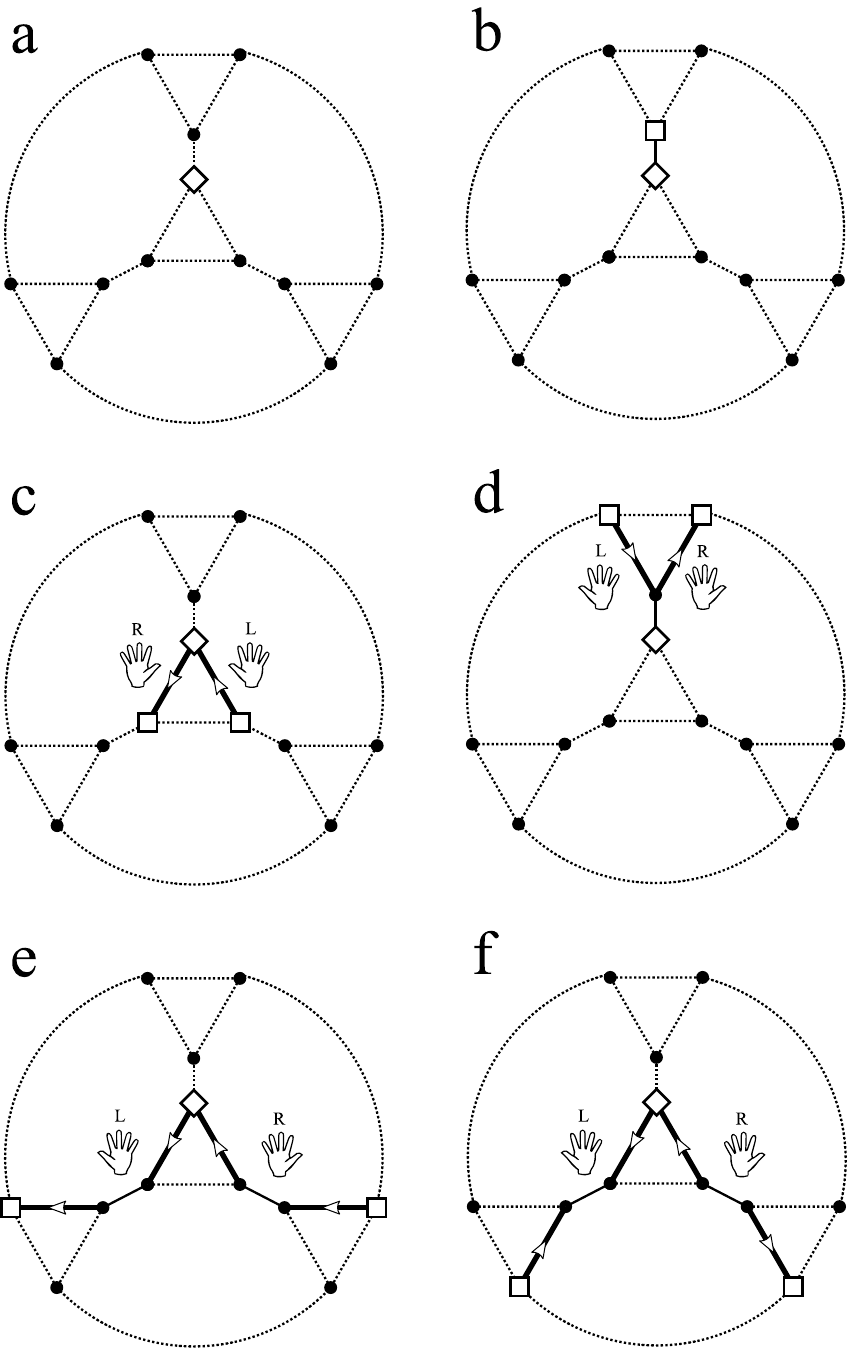}
\caption{Classes of interactions on the Thompson tetrahedron. Different classes of paths along the graph denote different slip system interactions. The first slip system of a given interaction is marked by a diamond, the second by a square. Shown are a) self, b) cross-slip, c)coplanar, d) glissile junction, e) Lomer-type, and f) Hirth-type interactions. Right- and left-handed interactions are marked.\label{fig:paths}}
\end{figure}

For a more intuitive representation of these conditions, please refer to Fig. \ref{fig:paths}, which shows these conditions represented as the various unique classes of paths one can take on the slip system graph structure shown in Fig. \ref{fig:key}. The classes and their right- and left-handed variants are shown as well. From these graphical representations, it is clear that the classes are not only defined by the types of connections traversed between two slip systems, but also the direction of the edges traversed. The former results in the traditional classification scheme, while it is the directionality of the graph edges which results in the handedness discussed above. For completeness, a table of the interactions of all 144 combinations of slip system is shown in Table \ref{tab:enum}, with left-handed interactions highlighted.

\begin{table}[]
\caption{Enumeration of all interaction types.\label{tab:enum}}
    \label{tab:my_label}
    \includegraphics[width = 0.7\columnwidth]{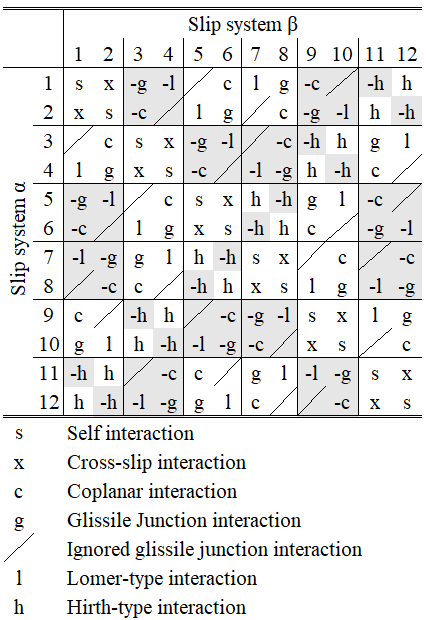}
    
\end{table}
    
 \subsubsection{Equivalence Spaces}
The handedness is central to the classification problem that presents itself to us in the analysis of correlation functions because it allows us to define coordinate transformations which map the Burgers vectors and dihedral planes onto constant directions. There are two such coordinate transformations $\rot$. In the case where the two interacting slip systems share a slip plane normal, the rotation is given by
\begin{equation}
\rot:=\left(\; \bb^{[\alpha]} \;|\; \chi (\bn^{[\alpha]} \times \bn^{[\beta]})\;|\; \bn^{[\alpha]} \;\right)^T.
\end{equation}
In the case where the two slip systems have different slip plane normal vectors, the rotation is given by
\begin{align}
    &\rot := \nonumber \\  & \left(\; (\bn^{[\alpha]} \wedge \bn^{[\beta]}) \times \bn^{[\beta]} \;|\; \chi (\bn^{[\alpha]} \wedge \bn^{[\beta]})\;|\; \bn^{[\beta]} \;\right)^T.
\end{align}

The result of these transformations is that for self (assumed right-handed) and coplanar interactions, the Burgers vector of the first (or only) interacting slip system is aligned to the x-axis and the shared normal direction to the z-axis. The y-axis is then defined in a right- or left-handed sense depending on the chirality of the interaction. For the remaining interactions—cross-slip (assumed right-handed), glissile junction, Lomer-type, and Hirth-type—the dihedral direction of the two planes is aligned to the y-axis and the second slip plane normal to the z-axis. The x-axis is then defined in a right- or left-handed sense according to the chirality of the interaction. 

These all represent simple coordinate rotations, although according to the interaction handedness these will be proper or improper rotations. Proper or improper, they map the dihedral planes and Burgers vectors onto constant directions in what will for the remainder of the work be referred to as the equivalence space of the respective class of interactions.

\subsubsection{Burgers reversals}
Lastly, we also define a sign by which the Burgers vector and line direction of the two slip systems ought to be reversed in order to make use of the above spatial rotations. Like the rotation tensors, the behavior is dependent on the two slip systems sharing or not their slip plane normal. In the case where the slip plane normal is shared, the second sign ought to be reversed according to the sign product of the two Burgers vectors:
\begin{align}
s^{[\alpha]}&=1 \nonumber\\
s^{[\beta]} &= b^{[\alpha]} \cdot b^{[\beta]}
\end{align}

In the case where the slip plane normal is not shared, both signs are reversed according to the sign product of the respective Burgers vector with the dihedral direction:
\begin{align}
s^{[\alpha]} &= b^{[\alpha]} \cdot \left( n^{[\alpha]} \wedge n^{[\beta]}\right)\\
s^{[\beta]} &= b^{[\beta]} \cdot \left( n^{[\alpha]} \wedge n^{[\beta]}\right)
\end{align}

\subsection{Final remarks regarding classification scheme}
In closing this section of the analysis, we show how these equivalence classes will be leveraged in the analysis of the two-point density data. Whereas previously we were considering the behavior of the product density,
\begin{equation}
    \brho\sab(\br,\br+\bdr;L),
\end{equation}
we will now be considering the qualitatively equivalent behavior of distinct groups of interactions. That is, for $[\alpha,\beta]\in \mathrm{type}$, where $\mathrm{type}$ is some class of interactions as outlined above, we now consider the behavior of the following product density:
\begin{align}
    \brho\TYPE(\brt,&\brt+\bdrt;L) \nonumber\\
    &\sim s\sa s\sb \brho\sab \left(\br,\br+(\rot)^{-1} \bdrt; L\right),
\end{align}
where $\brt$ here denotes a vector in the equivalence space and we have flipped both dislocation vectors $\brho^{[\alpha]},\brho^{[\alpha]}$ by the Burgers signs $s^{[\alpha]},s^{[\beta]}$.

In this manner, a given interaction class now can be treated \textit{in toto} by revising our definition of the equivalent pairs [Eq. (\ref{eq:pairs}])] to include all pairs which share a separation vector in the equivalence space:
\begin{equation}
    \Pi\TYPE(\bdrt) := \bigcup_{[\alpha,\beta]\in\mathrm{type}} \Pi\sab\left((\rot)^{-1}\bdrt \right).
\end{equation}
With this classification scheme in hand, we have reduced the number of unique correlation tensors which need to be calculated from 144 to 6.

\section{Calculation of Correlation Functions}
The current correlation calculations are performed using the same library of discrete dislocation dynamics simulations used in \cite{Anderson2021}. In summary, this represents 45 unique simulations of dislocations in copper which were run to 0.3\% plastic strain in uniaxial ([100] loading axis) strain-controlled simulations using the microMegas code \cite{Devincre2011}. {Because this code is lattice-based,  the lines were resampled to the midpoints of the discrete segments in order to fully populate the angular space \cite{Anderson2023}}. The dislocation densities in these simulations evolved from an initial value of 2 {\textmu}m$^{-2}$ to an average final value of 6 {\textmu}m$^{-2}$ and the simulation domain measured 4.40 x 4.87 x 5.74 {\textmu}m.

We have made use of the cloud-in-cell function \cite{Birdsall1969} as our choice of weight function for averaging the dislocations:
\begin{equation}\label{eq:CIC}
    w_L(\br):=\begin{cases}
        \prod_{i=1}^3 \left(1- \frac{|r_i|}{L}\right) & \text{for } |r_i| < L\\
        0 & \text{otherwise}
    \end{cases}
\end{equation}

For the details of the analytical line integral of this weight function over the straight segments used in discrete dislocation dynamics simulations, please refer to appendix 2. The main result is that for a collection of line segments $\lambda_i$ with line tangents $\bt_i$ passing through the cube $\Omega_L$ of side length $2L$ surrounding the point $\br$, the line integral of each can be evaluated as a weight $w_L(\lambda_i)$. This allows us to calculate both a vector-valued dislocation density and a scalar line density as [cf. Eq. (\ref{eq:CGrho})]:
\begin{align}
    \brho(\br;L) &= \sum_{\lambda_i}\bt_i w_L(\lambda_i),\\
    \rho(\br;L) &= \sum_{\lambda_i}w_L(\lambda_i).
\end{align}
These may not be of equivalent magnitude for volumes containing more than one dislocation line. The line bundle approximation only requires that the points where $|\brho|$ is not approximately equal to $\rho$ are of little interest.

The majority of the density calculations performed utilized a `discrete’ mesh of 720 cubic voxels along the longest box dimension. The fundamental length $L^*$ was taken as equivalent to this mesh spacing, 8.1 nm. The continuum density field calculations were performed using coarse-graining lengths L that were multiples of the mesh spacing, from 5 times the mesh spacing (40.5 nm) to 10 times this spacing (81 nm). A small minority of the calculations were performed comparing the continuum coarse-graining length of 81 nm and varying the fundamental discrete length $L^*$ from 16.2 nm down to 8.1 nm by increasing the number of mesh elements along the longest box dimension. 

Having density fields in hand, we now turn our attention to how these are used in the scheme discussed above to calculate correlation functions. As outlined above, once grouped into spatial equivalence classes, for every separation distance $\bdrt$, we have the scattered data:
\begin{align}
    \Big\{ \Big( D\TYPE_{ij}(\br,\br'),\; & C\TYPE_{ij}(\br,\br')  \Big) \nonumber \\ 
    &\forall \;(\br,\br') \in \Pi\TYPE(\bdr)\Big\}, \label{eq:Scatter_data}
\end{align}
where we have relabeled the discrete product density $D\TYPE_{ij}(\br,\br'):=\rho\TYPE_{ij}(\br,\br';L^*)$ and continuum product density $C\TYPE_{ij}(\br,\br'):=\rho\TYPE_{ij}(\br,\br';L)$ for notational clarity. An investigation of the correlation amounts to, for every separation distance, fitting the following linear relationship between the two:
\begin{equation}
    D\TYPE_{ij}(\bdrt) := C\TYPE_{ij}(\bdrt)g\TYPEij(\bdrt),
\end{equation}
where $g\TYPEij(\bdrt):= 1+ d\TYPEij(\bdrt)$ is a more compact relationship for this slope. {Notice, this is simply a compact notation for Eqs. (\ref{eq:17},\ref{eq:corr_def}), the definition of the correlation. This compact notation represents the data-driven context in which we will evaluate the correlation functions.}

For this reason, we outline a series of fitting methods for $g\TYPEij(\bdrt)$. These include: ordinary least squares fitting with discrete (OLSD) and continuum residuals (OLSC), and a principal component analysis of the covariance matrix (PC). All three methods make use of the covariance matrix:
\begin{equation}
    \mathrm{cov}(D,C,\bdrt) := \left( \begin{matrix}
        \frac{\ens{D^2}(\bdrt)}{D^2_\infty} & \frac{\ens{CD}(\bdrt)}{CD_\infty} \\
        \frac{\ens{CD}(\bdrt)}{CD_\infty} & \frac{\ens{C^2}(\bdrt)}{C^2_\infty}
    \end{matrix}  \right) 
\end{equation}
where the averages are now understood as empirical averages over the set $\Pi\TYPE(\bdrt)$:
\begin{equation}
    \ens{f}(\bdrt) =\frac{1}{|\Pi\TYPE(\bdrt)|}\sum_{(\br,\br')\in\Pi\TYPE(\bdrt)}f(\br,\br')
\end{equation}
Here, $|\Pi\TYPE(\bdrt)|$ represents the number of pairs in the set $\Pi\TYPE(\bdrt)$. Additionally, each average is measured relative to the uncorrelated [in the sense of Eq. (\ref{eq:uncorrelated})] average. That is, for some function $h(C,D) := h_1(C)h_2(D)$ satisfying the requirements of Eq. (\ref{eq:uncorrelated}), we may express the uncorrelated average as
\begin{widetext}
\begin{align}
    h_\infty(C\TYPE_{ij},D\TYPE_{ij}) = &\left[ \frac{1}{\sum_{(\alpha,\beta)\in \mathrm{type}} |S\sa|} \sum_{(\alpha,\beta)\in \mathrm{type}} \sum_{\br\in S\sa} h_1\big(\rho\sa_i(\br;L)\big) h_2\big(\rho\sa_i(\br;L^*)\big)   \right] \nonumber \\
    & \times \left[ \frac{1}{\sum_{(\alpha,\beta)\in \mathrm{type}} |S\sb|} \sum_{(\alpha,\beta)\in \mathrm{type}} \sum_{\br\in S\sb} h_1\big(\rho\sb_j(\br;L)\big) h_2\big(\rho\sb_j(\br;L^*)\big)   \right].
\end{align}
\end{widetext}

Each of the four fitting schemes outlined above utilize the covariance averages, as follows:
\begin{align}
    g_{\mathrm{OLSC}}(\bdrt)&:= \frac{\ens{D^2}(\bdrt)}{\ens{CD}(\bdrt)} \frac{CD_\infty}{D^2_\infty}\label{eq:corr_OLSC}\\
    g_{\mathrm{OLSD}}(\bdrt)&:= \frac{\ens{CD}(\bdrt)}{\ens{C^2}(\bdrt)} \frac{C^2_\infty}{CD_\infty}\\
    g_{\mathrm{PC}}(\bdrt) &:= \frac{\beta_D^{(1)}}{\beta_C^{(1)}} \label{eq:corr_PC}
\end{align}
where $\boldsymbol{\beta}^{(1)}:=(\beta_D^{(1)},\beta_C^{(1)})$ represents the eigenvector of the covariance matrix corresponding to the highest eigenvalue. This expansion is performed as follows:
\begin{equation}
    \mathrm{cov}(D,C):=(\boldsymbol{\beta}^{(1)}|\boldsymbol{\beta}^{(2)})\cdot\left(\begin{matrix}
        \sigma_{(1)}^2 &0\\ 0 &\sigma_{(2)}^2
    \end{matrix}\right) \cdot (\boldsymbol{\beta}^{(1)}|\boldsymbol{\beta}^{(2)})^T
\end{equation}
with $\sigma_{(1)}^2 \geq \sigma_{(2)}^2$. That is, the covariance matrix is diagonalizable into two statistically uncorrelated linear combinations of $C$ and $D$. The eigenvector corresponding to the combination with the larger variance is chosen as a correlation function [Eq. (\ref{eq:corr_PC})]. As an additional measure of the variance in the scatter relationship [Eq. (\ref{eq:Scatter_data})] which is not captured by the linear correlation relationship, we will also examine the following relation as the fluctuation variance:
\begin{equation}
    \sigma_\mathrm{f}^2 = \frac{\sigma_{(2)}^2}{\sigma_{(1)}^2+\sigma_{(2)}^2}. \label{eq:fluc_var}
\end{equation}
{The relative strength of the principal component analysis just described is that the random components  of the data along each principal component (the correlation fit line and the remaining fluctuation variance) are statistically uncorrelated. For a more detailed description of principal component analysis, cf. \cite{Shlens2003}.}

\section{Results}
\begin{figure*}
    \includegraphics[width=172mm]{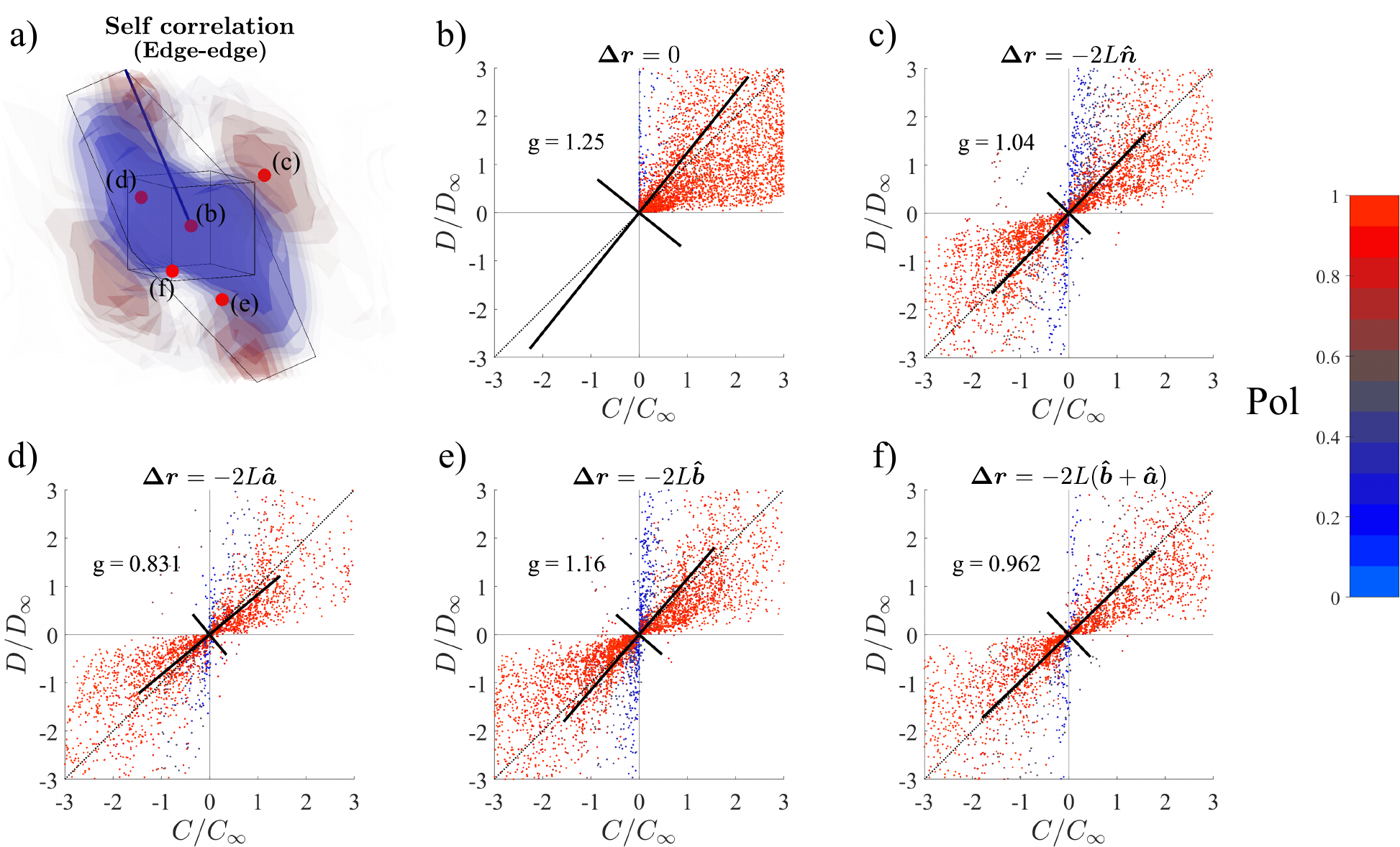}
    \caption{Scatter plots of discrete and continuum density product data. (a) Volumetric representation of the edge-edge self-correlation function given for reference of separation space locations. {The slip plane (hexagon), Burgers direction (blue line), and coarse-graining volume (cube) are shown for reference. In this figure, blue regions are anti-correlated regions ($d<1$), and red regions are positively correlated ($d>1$). } (b-f) Scatter plots of the underlying discrete and continuum density products, and the linear relationship which the correlation function describes. In (a), blueregions correspond to anti-correlated regions ($g<1$) and red to correlated regions ($g>1)$), while transparent regions correspond to uncorrelated regions ($g=1$). Orienting features of (a) include the Burgers vector direction (blueline), slip plane (hexagon), and the coarse-graining volume (cube). The points in separation space examined in (b-f) are shown as well. In (b-f), the dotted line represents the relation that would obtain in the perfectly uncorrelated case (a slope of unity). The solid lines show the principal components of the covariance matrix. The slope of the larger component is taken as the correlation value and reported in each plot. Points in (b-f) are colored according to the effective polarization [Eq. (\ref{eq:efPol})], with blue denoting low polarization ($\ll 1$) and red denoting high polarization ($\approx 1$) according to the colorbar shown on the right. \label{fig:Scatter}}
\end{figure*}

We begin this section with a figure showing the scattered nature of the product density data (Fig. \ref{fig:Scatter}) as an aid for conceptually understanding the correlation fitting procedure. The first plot shows a spatial correlation function with several points labeled. Corresponding to each of these points in the separation space, the product density at the two scales is represented as scattered data [cf. Eq. (\ref{eq:Scatter_data})]. 
The linear relationship passes through the origin--this is required due to the lack of constant terms in Eq. (\ref{eq:corr_def})\footnote{The lack of constant terms in the definition of the correlation relation is no accident. Rather, it is a corollary of a certain property of the coarse grained densities in line bundle ensembles, namely that the coarse densities are absolutely continuous with respect to the discrete densities \cite{Anderson2021}.}--but differs from point to point; this results in a correlation field that is a function of separation distance (cf. Fig. \ref{fig:Scatter}a). 

An entirely uncorrelated product density would result in the scattered data falling perfectly on the line of unit slope passing through the origin. The principal components of the covariance matrix for the scattered data at each point are shown in Fig. \ref{fig:Scatter}(b-f) as black lines, the length of which denotes the variance along each component. The direction of the longer component gives the PC fit for the correlation function [Eq. (\ref{eq:corr_PC})], which is reported in each plot. This direction also defines the fluctuation direction (these two directions are orthogonal), and the length of the shorter component is related to the fluctuation variance [Eq. (\ref{eq:fluc_var})]. Thus, Fig. \ref{fig:Scatter} not only shows qualitatively what the correlation represents, but also the limits of the information it contains. This limitation, quantified by the fluctuation variance, will be revisited at the end of the present section and we will examine its implications in a later section. 

One additional feature of the scatter plots in Fig.\ref{fig:Scatter}(b-f) is that the the effective polarization of each data point is shown. That is,
\begin{equation}
    \mathrm{Pol} = \sqrt{\frac{|\brho (\br)|}{\rho(\br)}\frac{|\brho (\br')|}{\rho(\br')}}. \label{eq:efPol}
\end{equation}
It is worth noting that the linear relationship is especially erroneous for data points with low effective polarization; such points have high discrete product density but low continuum product density and cluster along the vertical line that would represent an infinite value of the correlation. Such data obviously violate the line bundle assumption; however, we note that they represent a negligibly small portion of the total data.  

\begin{figure*}
    \includegraphics[width=129mm]{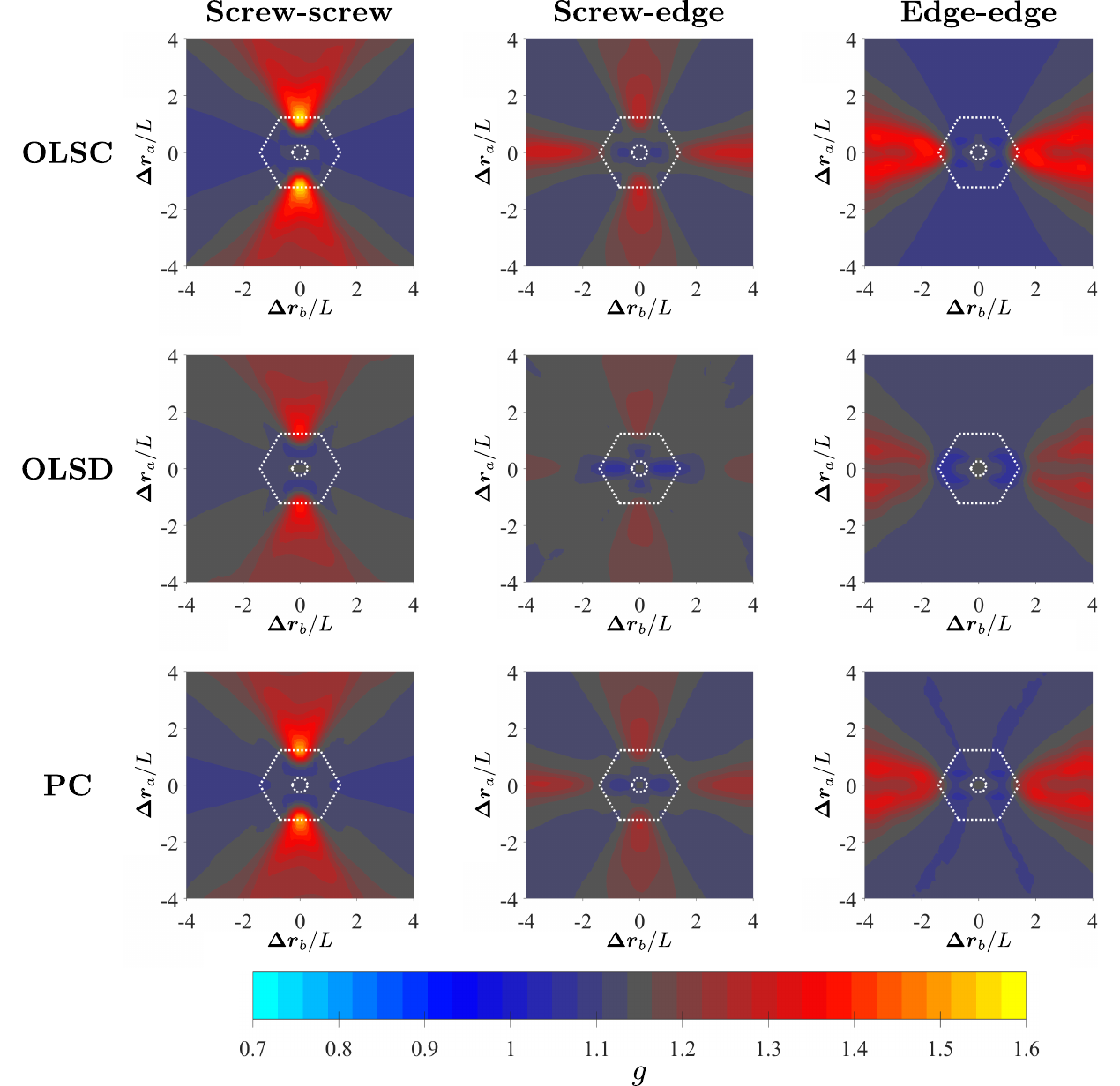}
    \caption{Various fits for the self-correlation function in the slip plane. Self-correlation functions are shown for separation vectors in the slip plane, where $\Delta r_b$ and $\Delta r_a$ are the separation in the Burgers vector direction and the line direction of an edge dislocation, respectively. The three methods of fitting the correlation are compared, namely  ordinary least-squares with continuum (OLSC) and discrete (OLSD) residuals, and the prinicpal component (PC, cf. Fig. \ref{fig:Scatter}). These are shown for the screw-screw, screw-edge, and edge-screw components of the self-correlation function. The morphology of the correlation function is unaffected by the choice of fitting method, although the magnitude of the respective maxima differs slightly between the three methods. The faint hexagons represent the boundaries of the fundamental and coarse-graining volumes (cube of side length $2L^*$ and $2L$, respectively). \label{fig:Fits}}
\end{figure*}

Fig. \ref{fig:Fits} shows a typical self-correlation field (self in the sense defined in section IV, as pertaining to density products of like slip system). Results for the various fitting methods outlined above are shown for separation distances which lie in the slip plane, i.e. those separation vectors which have no component in the slip plane normal direction. The three unique tensor components of the self-correlation field (screw-edge and edge-screw are symmetric under inversion of separation space) are shown for the three fitting methods described in Eqs. (\ref{eq:corr_OLSC}-\ref{eq:corr_PC}). The distinctive morphologies of the self-correlation field can be seen in this figure. All of the fitting methods demonstrate roughly equivalent morphologies, differing mainly in the magnitude of their respective extrema. The principal component analysis fitting method, however, will be utilized in the remainder of the manuscript because of the property that the error axis is statistically uncorrelated (in the sense of null covariance) to the correlation axis. This fluctuation axis, briefly discussed above, will be the subject of later discussion. 

\begin{figure*}
    \includegraphics[width=172mm]{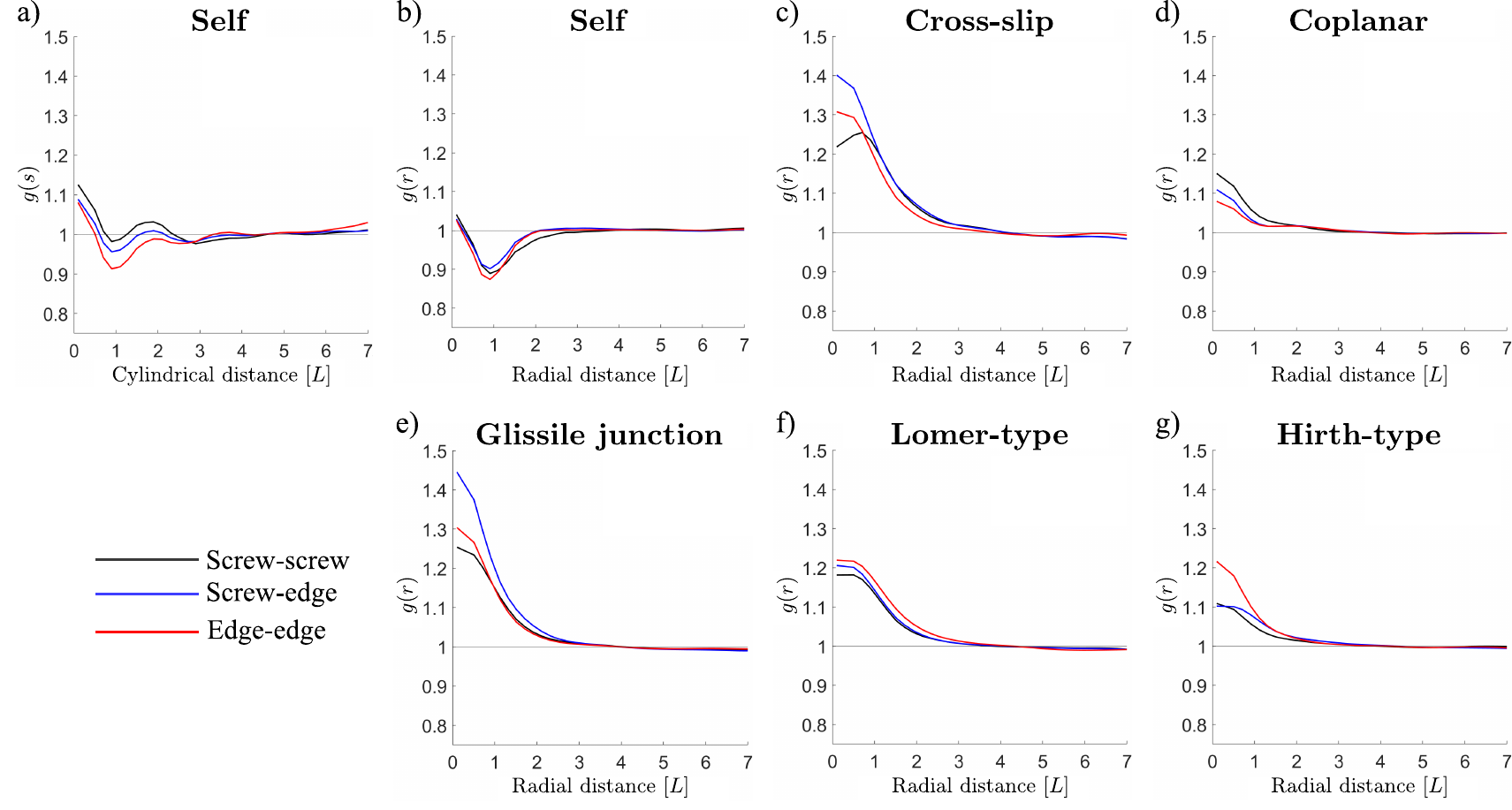}
    \caption{Radial plots for all correlation functions. (a) Cylindrical correlation function [Eq. (\ref{eq:cyl_corr})] for the self-correlation . (b-g) Spherical correlation function [Eq. (\ref{eq:spher_corr})] from the correlation functions corresponding to each interaction class. \label{fig:radcorr}}
\end{figure*}
\begin{figure*}
    \includegraphics[width=129mm]{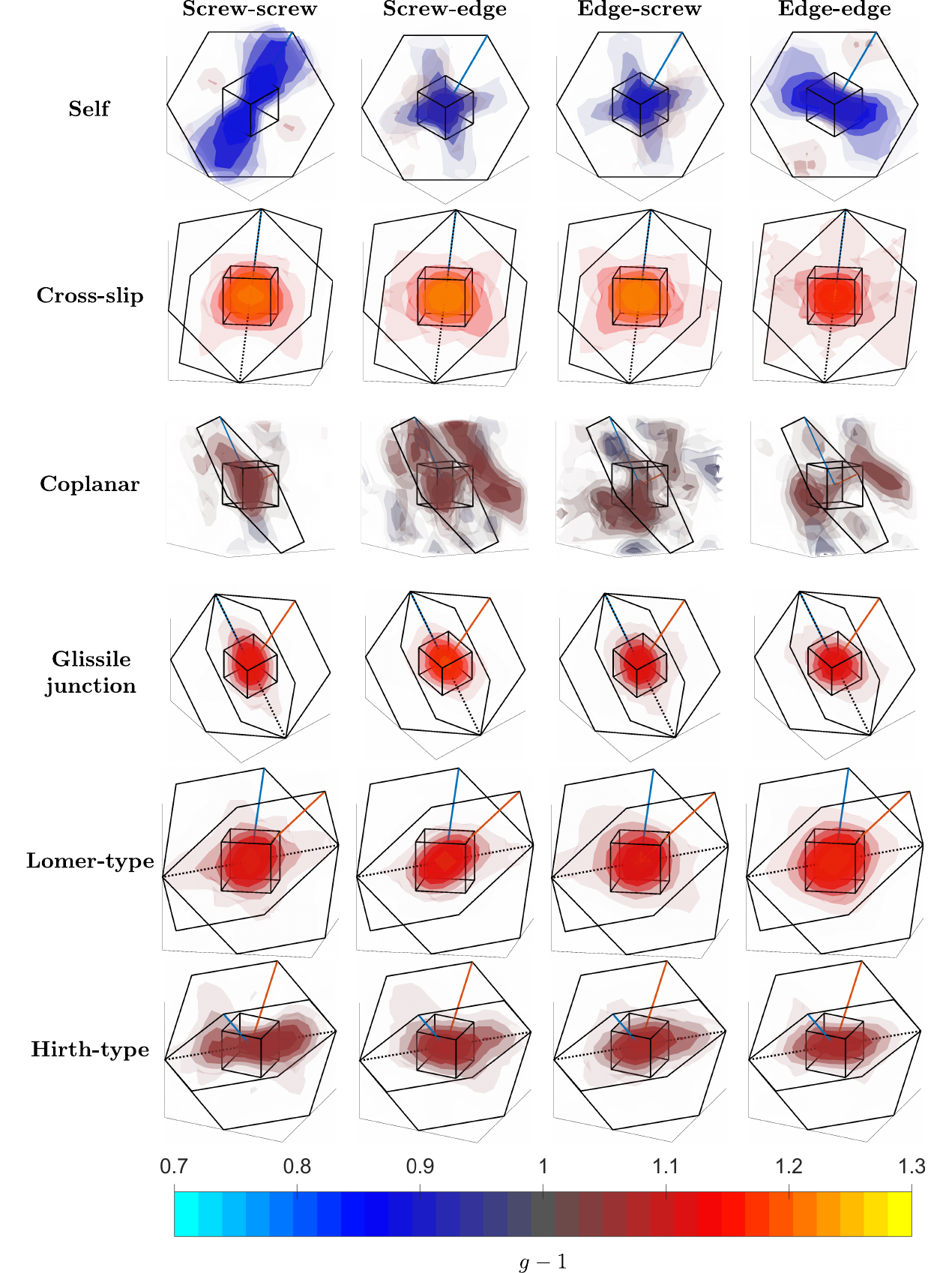}
    \caption{Volumetric plots of all correlation functions. Volumetric isosurfaces of the correlation functions for all interaction classes are colored according to the value of the correlation function. The color scale, denoting the value of the correlation, is constant across all plots. The transparency scale of each plot ranges from the maximum deviation from unit correlation (opaque) to uncorrelated (unit correlation, transparent). For spatial reference, several relevant spatial features are shown. Cubes show the coarse-graining volume (cube of side length $2L$). Hexagons show the intersection of the slip planes of the respective slip system with the cube of side length $6L$. Dotted line shows the intersection of the two slip planes if applicable. blueand orange lines show the Burgers vector of the first and second slip systems, respectively. \label{fig:volume}}
\end{figure*}

Next we move on to the full cross-correlation results for each type of density field combination enumerated in section IV. Reporting on the cross-correlation is one of the main objectives of the present work. Firstly, in Fig. \ref{fig:radcorr} we show the radial convergence of these cross-correlations. The spherical correlation functions shown in Fig. \ref{fig:radcorr}(b-g) are the average correlation value over the spherical shell of radius r:
\begin{equation}
    g(r) := \frac{1}{4\pi r^2} \int_0^{4\pi} r^2d\Omega \; g(\br) .  \label{eq:spher_corr}
\end{equation}
Because of the strong anisotropy of the self-correlation function surrounding the slip plane, we also compute for that correlation function a cylindrical correlation function. This is computed as the average over the disc of height $L$ and radius $s$ centered on the origin:
\begin{equation}
    g(s):=\frac{1}{2\pi sL} \int_{-L/2}^{L/2}dr_n \int_0^{2\pi} d\varphi \; g(\br).\label{eq:cyl_corr}
\end{equation}
In addition to these radial correlation functions, full volumetric contour plots are also shown in Fig. \ref{fig:volume} to display the full spatial morphology of the correlation functions. 

Of foremost significance to the dislocation system is the range of the correlation functions, which is displayed by the radial convergence shown in Fig. \ref{fig:radcorr}. As these ranges define the degree of non-locality which must be considered for corrections to the Peach-Koehler functional, it is reassuring that these converge to uncorrelated values in most cases by 2-4 times the coarse-graining length. Of particular prominence are the longer-ranged and more strongly correlated cross-slip and glissile junction type correlations. Of least prominence are the coplanar correlations, which densities it would seem are almost completely uncorrelated in the simulations analyzed. 

Additional insights into the features of the correlation functions can be gained from the volumetric renderings. In Fig. \ref{fig:volume}, it is clear that at least some degree of anisotropy is evident in all cross-correlations. The self-correlation is seen to be strongly anisotropic, with significant variation in the slip plane (cf. Fig. \ref{fig:Fits}) as well as strong anti-correlations ($g<1$) for small distances normal to the slip plane. The dominance of these anticorrelated regions just off the slip plane can be seen in the cylindrical and radial correlations of the self-interaction type (Fig. \ref{fig:radcorr}a-b). While not as dramatic as in the self-correlations, the cross-correlations do display some weak anisotropies. Notably, they seem to decay more slowly  in the direction of the dihedral between their two slip planes.

\begin{figure*}
    \includegraphics[width=129mm]{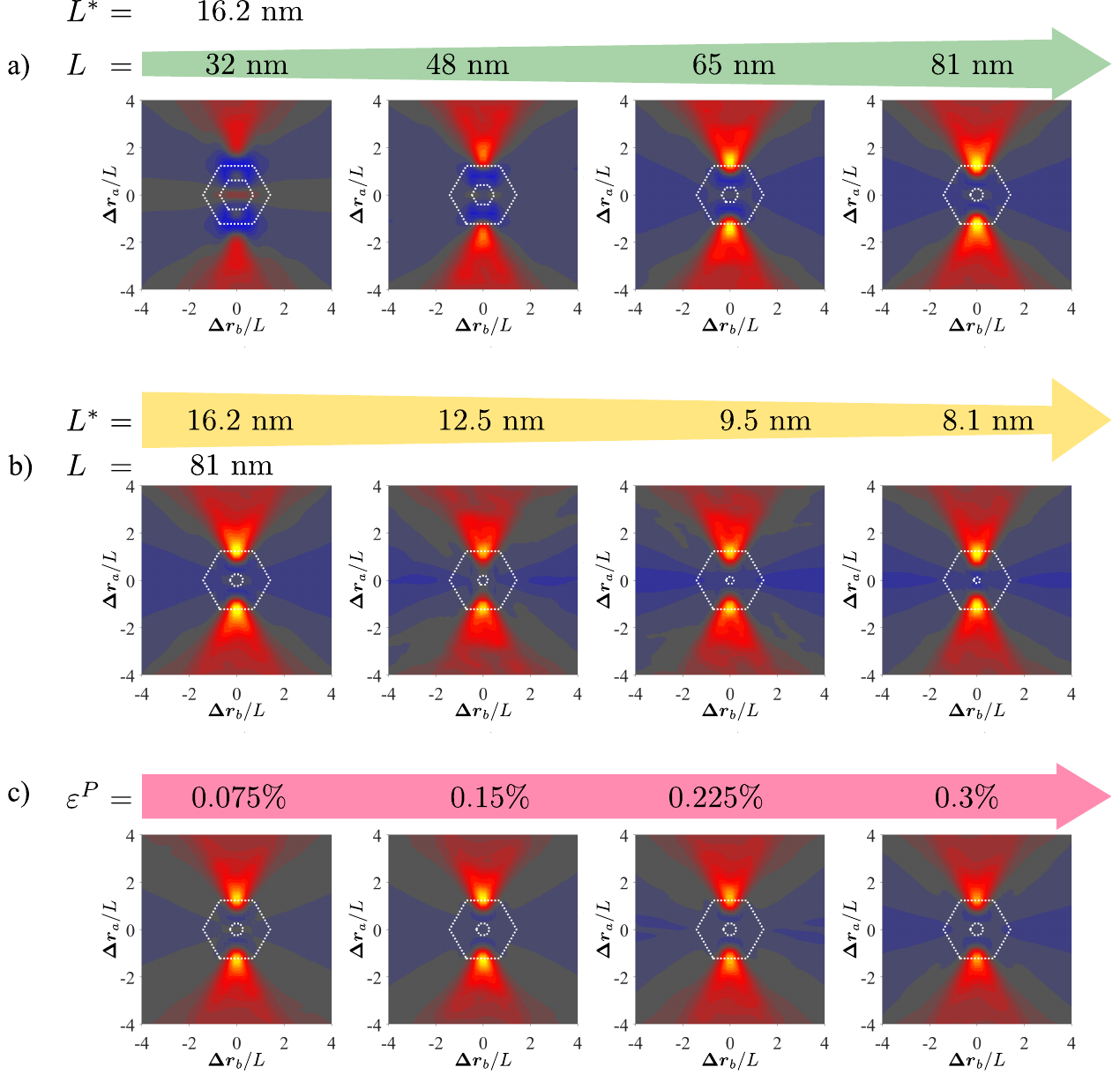}
    \caption{Dependencies of screw-screw self-correlation on various calculation parameters. (a) Self-correlations calculated with various coarse-graining lengths and a fixed fundamental length. (b) Self-correlations calculated with various fundamental lengths and a fixed coarse-graining length. (c) Self-correlations calculated from configurations with various values of plastic strain. Note, all separation distances scaled by the coarse-graining length. \label{fig:deps}}
\end{figure*}

Of considerable interest to any implementation of the correlation functions is the question whether these functions vary as the system evolves, that is, as the total dislocation line length in the volume increases. Two additional parameters are the choices of fundamental and coarse-graining lengths, the length scales associated with the discrete and continuum representations of the dislocation densities. These remain fixed in a continuum theory, but in principle they should affect the correlation functions. Screw-screw self-correlation calculations which independently vary each of these parameters are shown in Fig. \ref{fig:deps}. The significant dependencies are on the relationship between the discrete and continuum coarse-graining lengths. These are seen in the first two series of plots [Fig. \ref{fig:deps}(a-b)].  In the former (a), correlation functions are calculated from density fields with 16 nm fundamental length and coarse-graining lengths increases from 32 to 81 nm. In the latter (b), the coarse-graining length is held fixed at 81 nm while the fundamental length is decreased from 16 to 8 nm. From these plots, it can be seen that the relevant spatial features of the self-correlation functions are relative to the discrete and continuum coarse-graining lengths. This strong dependence demonstrates that the correlations arise as features of the spatial averaging process itself, rather than some particularly favored dislocation arrangements. Additionally, the dependence of the correlation function on plastic strain is shown in Fig. \ref{fig:deps}(c). The correlation functions are seen to be stable with respect to strain, showing little evolution over the course of the simulation. This relationship will be further elaborated in the discussion section to follow.

\begin{figure*}
    \includegraphics[width=129mm]{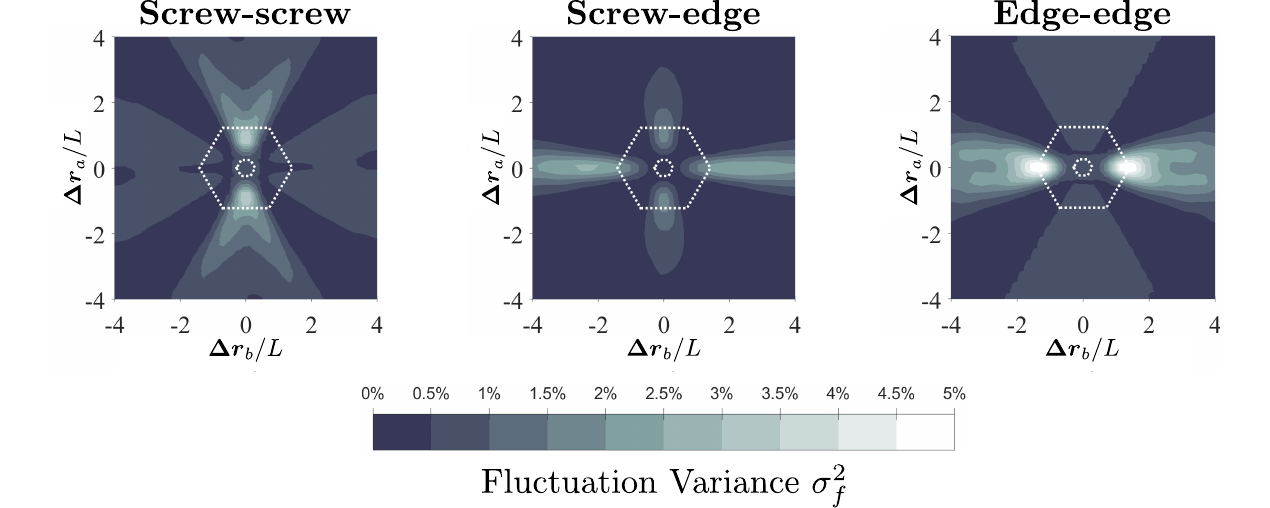}
    \caption{Fluctuation variances of the self-correlation calculation. Displayed are the fluctuation variance field for the self-correlation calculation as a fraction of the total variance in the product density data at each point [Eq. (\ref{eq:fluc_var})] for separation vectors in the slip plane.  \label{fig:variance}}
\end{figure*}

Lastly, the portion of the variance at each separation distance which was unexplained by the correlation function (cf. Fig. \ref{fig:Scatter}) is shown in Fig. \ref{fig:variance}. While the correlation was defined to capture the error which is introduced in the discrete product density by naively assuming it to be equal to the continuum product density, some error remains even after defining a correlation function. This is due to the closure of the density hierarchy at order two as well as the suppression of the fully six-dimensional dependence of the correlation function. Nonetheless, the noticeable spread off the correlation axis in the scattered data shown in Fig. \ref{fig:Scatter} still only accounts for no more than 5\% of the total variance at any separation distance. Of note, however, is that the areas with the highest fluctuation variance correspond to the areas of strong correlation ($g>1$). These regions, occurring near the boundary of the coarse-graining region, arise due to the small continuum field values which occur near the boundaries of the coarse-graining volume in  the ‘unconnected’ direction. 

\section{Discussion \label{sec:Discuss}}
{The calculated correlation functions reported in the previous section, while changing little with strain, depend strongly on the size of the coarse-graining volume. This suggests that correlations arise due to the averaging of a field of planar lines, rather than encoding a contingent arrangement of the microstructure. This is in contrast to theories at longer length scales, where the correlation begins to capture the length scale of such a microstructure  \cite{Groma1999}.} Nonetheless, the correlations measured above still encode important kinetic information. In this section we discuss our interpretation of what information that is by examining the impact of the presented results on future investigations in vector density continuum dislocation dynamics. In particular, we will examine the implications of these correlation functions for mean-field continuum dislocation dynamics. However, we also discuss some implications of our findings which gesture towards the possibility of stochastic theories of continuum dislocation motion as well.

\subsection{Correlation effects in mean-field continuum dislocation dynamics}
Current vector-density continuum dislocation dynamics simulations only treat the mean-field portion of the kinetic equations [Eqs. (\ref{eq:Kinetic_1}-\ref{eq:Kinetic_2})]. That is, they ignore the correlation force given by Eq. (\ref{eq:Kinetic_2}). The mean field force term is generally evaluated by considering the mechanical solution to the eigenstrain produced by the Kr{\"o}ner-Nye tensor (El-Azab 2018):
\begin{align}
    \nabla \cdot \boldsymbol{\sigma}(\br) &= 0, \label{eq:equil}\\
    \boldsymbol{\sigma}(\br) &:= \mathbb{C} \; : \; \left(\nabla \boldsymbol{u}_\mathrm{tot} - \boldsymbol{\beta}_\mathrm{p}(\br)\right)_\mathrm{sym}, \label{eq:Hooke}\\
       \boldsymbol{\beta}_\mathrm{p}(\br) &= \nabla \boldsymbol{z} (\br) + \boldsymbol{\chi}(\br), \label{eq:beta_def}\\
    \nabla \times \boldsymbol{\chi}(\br) &= \boldsymbol{\alpha}(\br),\label{eq:chi_def} \\
    \boldsymbol{\alpha}(\br) & := \sum_{\eta=1}^{N_S} \brho^{[\eta]}(\br) \otimes \bb^{[\eta]}. \label{eq:alpha_def}
\end{align}
These equations allow the total displacement field--and hence the stress field $\boldsymbol{\sigma}$--to be calculated from the dislocation arrangement. They do so in the following manner. The total displacement gradient is (additively) decomposed into an elastic part and a plastic part. These are distinguished by the constitutive relation [Hooke's law, Eq. (\ref{eq:Hooke})], whereby only the elastic distortion $\nabla \boldsymbol{u}_{\mathrm{tot}}- \boldsymbol{\beta}_\mathrm{p}$ is used to compute the stress field. The plastic distortion ($\boldsymbol{\beta}_\mathrm{p}$), of course, can be decomposed [Eq. (\ref{eq:beta_def})] into a compatible part--i.e. a part which can be represented by a gradient of some plastic displacement field $\boldsymbol{z}$--and an incompatible part $\boldsymbol{\chi}$ which has a nontrivial curl. The closure fault of this plastic distortion corresponds precisely to the Burgers circuit definition of dislocations, allowing the curl of the incompatible distortion to be represented in terms of the Kr{\"o}ner-Nye tensor $\boldsymbol{\alpha}$. The Kr{\"o}ner-Nye tensor, being a measure of the geometrically necessary dislocation content of the crystal region, is given straightforwardly in terms of the dislocation densities [Eq. (\ref{eq:alpha_def})]. In practice, the constitutive equations [Eqs. (\ref{eq:Hooke},\ref{eq:equil})] give a second order differential equation for the elastic displacements $\boldsymbol{u}_\mathrm{tot}-\boldsymbol{z}$ in terms of the dislocation arrangement. Once these displacements are fixed, Hooke's law [Eq. (\ref{eq:Hooke})] yields the stress field. This solution scheme, known as field dislocation mechanics (FDM) is commonly solved by finite element \cite{Lin2021} or spectral methods \cite{Brenner2014}.

A well-known but oft-minimized downfall of the FDM approach is that the short-range stress contribution is ignored, resulting in a dislocation transport equation which is, \textit{a priori}, incorrect \cite{Vivekanandan2022}. One creative approach to solve this problem is to utilize FDM within DDD simulations to solve for the long-range stress field while solving the short-range stresses and dislocation transport in terms of the discrete dislocation lines \cite{Bertin2019}. This may prove advantageous to DDD formalisms by reducing computational cost, but continuum formalisms are necessary to push into geometrically nonlinear regimes of strain \cite{Anderson2022,Starkey2020}. In continuum dislocation dynamics, a correlation dependent correction to the transport equation is necessary to capture the effects of short-range stresses. 

The specific form of the correlation force term $\Fsub{C}$ leverages the weak line bundle assumption to allow a straightforward calculation of the effective stress field needed to recapture the true dislocation dynamics. While the mean-field contribution to the mechanics is best calculated by FDM (with which it is possible to treat the nuances of geometric nonlinearity), the correlation portion can be treated using the infinitesimal stress kernel due to the short-range nature of the integral: the stress kernel converges as $O(r^{-1} )$ and the correlations presented here are of finite range. 

The degree of non-locality which needs to be considered in a model that evaluates the correlation effects can be gleaned from the radial convergence of the correlation functions (Fig. \ref{fig:radcorr}). As previously noted, these decay to an uncorrelated state over 2-4 coarse-graining lengths. We note that, by construction, the coarse-graining length represents the typical variation length of the density field. As a result, a local density approximation to the correlation force functional would be inappropriate. While it is possible that gradient theories may be considered to deal with this non-locality, further investigation is needed to understand what mathematical tools would be needed to evaluate the non-local correlation force integral.

Having briefly touched upon possible implementations of the correlation force, we now ask: what is gained by incorporation of the short-range stress effects? Energetically, the correlation introduces an alternative mechanism to store strain energy in local neighborhoods (i.e. at  small separations). Kinetically, this should result in the system preferring the coagulation of dislocation densities to minimize the more costly long-range strain energies. This shielding effect is thought to be the key to the onset of dislocation pattern formation in deformed metals. It is well known that the inclusion of \textit{a posteriori} phenomenological local energy storage terms (i.e. Taylor-type hardening terms) in CDD models leads to rudimentary pattern formation, although at low strains compared to experiment \cite{Vivekanandan2021,Xia2016,Sandfeld2015}. Our expectation is that inclusion of the short-range stress effects in the form of correlation dependent transport terms will naturally give rise to dislocation patterns without the need to resort to \textit{ad hoc} terms designed to force pattern formation \cite{Ispanovity2020}. 

\subsection{Stochastic effects}
A closure of the density hierarchy at any order will at some level mean projecting away the fluctuations of the higher order densities (i.e., higher than 2-point). These fluctuations are commonly ignored, but they do play a role in the true dynamics of the system, especially when the six-dimensional pair space is reduced to a three-dimensional separation space. As was immediately clear in the present investigation of the correlation functions (cf. Fig. 3), the linear relationship between the discrete and continuum density products $D = gC$ does not account for all of the variance in the value of D. In fact, as is seen from Figs. 3 and 8, a significant amount of fluctuation variance remains in the sense of Eq. (51). This variance could be incorporated into the model by means of a suitable stochastic process. In such a case, the stochastic process would be superadded to the correlation relationship as follows:
\begin{align}
    D\TYPE _{ij}(\br,\br')=& C\TYPE _{ij} (\br,\br') g\TYPEij(\br-\br') \nonumber \\
    &\;\; + X\TYPEij_s(\br-\br').
\end{align}
The exact stochastic process that would best suit this relationship is beyond the scope of this work. However, for heuristic purposes, we consider the following:
\begin{equation}
    X\TYPEij_s(\bdr) := \frac{1}{g\TYPEij(\bdr)} \varsigma(\bdr) B_s,
\end{equation}
where $B_s$ is some elementary stochastic process, say a Brownian motion. By the dependence on $g^{-1}$, this process is naturally uncorrelated to the correlation effects (via the principal component analysis). The standard deviation of this process $\varsigma$ can then be given a suitable functional form in terms of the distance along the correlation axis, as it appears from Fig. \ref{fig:Scatter} that there are scaling effects at play.

The above is not meant, by any means, to be a firmly held hypothesis. Rather it is simply a gesture at one possible direction of development one might pursue in order to consider the effects of higher than second order density fluctuations. If pursued, this line of reasoning would be well-suited to investigations in terms of established projection operator methods such as the GENERIC framework \cite{Ottinger1998}. This projection operator approach to the stochastic effects would differ in method but be similar in results to two-dimensional investigations into the density-based mobility of dislocations \cite{Kooiman2015}. That is, the higher-order effects could alter the drag law defining the relationship between effective Peach-Koehler force and the time derivative of the density field.

\section{Concluding Remarks}

In this work, we have given a full treatment of the coarse-grained kinetics proper to a vector-density based theory of continuum dislocation dynamics. The kinetic theory which we developed naturally gave rise to dislocation correlation functions, which were defined in a straightforward manner. Utilizing a novel, chiral classification of the FCC slip systems, we were able to show that the full number of correlation functions needed to describe the kinetics of the vector-densities is 24, corresponding to 6 interaction classes and interactions between 4 various combinations of vector-density components. All 24 of these correlation functions were then computed from discrete dislocation dynamics simulations. The same were seen to depend strongly on the coarse-graining length used to define the continuum vector-density, but only weakly on the cumulative  plastic strain and total dislocation density. This implies that the correlations in vector-density schemes primarily encode line connectivity  information rather than some favorable microstructural arrangement of dislocations.

The evaluation of the correlation functions unveiled several key features thereof. First, the dislocation correlations were found to be of moderate range on the order of 2-4 times the coarse-graining lengths. The self-correlations, corresponding to the interaction of dislocations of like slip system, were seen to be more strongly anisotropic than the correlation of dislocations of unlike slip systems. Of these cross-correlations, the cross-slip type interactions were seen to have the strongest correlations, while the coplanar type interactions showed negligible correlation.

The implications of the present work for future progress in vector-density continuum dislocation dynamics are profound. In discussion of these implications we especially note the effect of the correlation functions on driving forces in the transport equations, where it is expected that these correlations can open a mechanism for dislocation pattern formation in deformed metals. Additionally, we discussed the stochastic fluctuations not captured in the correlation function which the present investigation brought to light and how these might lead in future investigations to a coarse-grained mobility function.

In summary, we have presented here the complete correlation information necessary to compute the average evolution of the dislocation density field. The full set of dislocation correlation functions now lies at our fingertips and the short-range interactions of dislocations can now be recovered in vector-density continuum dislocation dynamics. With this new tool, we have scaled one more barrier in our understanding of dislocations systems; what new plastic phenomena may now lie within our sight?

\section*{Appendix}
\subsection{1. Form of the Peach-Koehler functional}
In discrete dislocation dynamics, the velocity law used is normally one of overdamped motion. That is, there is some force field $\boldsymbol{f}$ which acts on a dislocation, and the velocity simply linearly related to that force:
\begin{equation}
    B\boldsymbol{v}(\br) := \boldsymbol{f}(\br),
\end{equation}
where $B$ is a viscous drag coefficient. In face-centered cubic crystals, a tensorial factor of $(\mathbb{I} - \bn\sa\otimes\bn\sa)$ is applied to project the force onto the slip plane \cite{Sills2016a}, but this will be omitted for brevity. 

Very early in the development of dislocation theory, it was realized that the configurational force which acts on the dislocation line--the negative of the work done by moving a dislocation some small distance--is given by the famous Peach-Koehler force [7]:
\begin{align}
    \boldsymbol{f}(\br)&:= \left(\bb\sa\cdot \boldsymbol{\sigma} (\br) \right) \times \bxi(\br)\\
    &= \left[\bb\sa \cdot \left(\boldsymbol{\sigma}^0 + \boldsymbol{\sigma}^\mathrm{(int)} \right)\right]\times \bxi\sa,
\end{align}
where we have introduced the stress field $\boldsymbol{\sigma}$ and its decomposition into a homogenous remote stress $\boldsymbol{\sigma}^0$ and the internal stress field $\boldsymbol{\sigma}^\mathrm{(int)}$ arising from the other dislocations in the system.

In an isotropic elastic medium, the internal stress field can be expressed as an integral with respect to the discrete dislocation density \cite{Hirth1968}. That is, the internal stress field is a functional of the discrete dislocation density:
\begin{equation}
    \sigma^\mathrm{(int)}_{ij} = \sum_{\beta=1}^{N_S} \int_\mathcal{M} \hat{\sigma}\sb_{ijk} (\br-\br')\sb \varrho\sb_k(\br') d^3 \br'
\end{equation}
where $\hat{\sigma}\sb_{ijk}$ represents the third rank Green's function for the stress field produced by a dislocation with Burgers vector $\bb\sb$. This Green's function is given by 
\begin{align}
    \frac{4\pi}{\mu} \hat{\sigma}_{ijk}\sb(\br) &= \frac{1}{2}\left( s\sb_i(\br) \delta_{jk} + s\sb_j(\br) \delta_{ik} \right) + \frac{1}{1-\nu} S\sb_{ijk}(\br)\\
    s\sb_i(\br) &:= (\bb\sb\times\nabla)_i \left(\frac{1}{|\br|}\right)\\
    S\sb_{ijk}(\br) &:= (\bb\sb\times\nabla)_k \left(G_{ij}-\frac{2\delta_{ij}}{|\br|}\right)\\
    G_{ij}(\br) &:=(\nabla \otimes\nabla)|\br| = \frac{\delta_{ij}}{|\br|}-\frac{r_i r_j}{|\br|^3}
\end{align}
It can be seen that the stress kernel itself is independent of any line information. Thus, it suffices to express the of Peach-Koehler force as a vector valued functional of the line information as follows:
\begin{widetext}
    \begin{equation}
        \left(\FPK\left[\bxi\otimes\bdrho\sb(\br') \right] \right)_l := \epsilon_{jkl} b\sa_i \xi\sa_k(\br) \left( \sigma_{ij}^0 + \sum_{\beta=1}^{N_S} \int_\mathcal{M} \hat{\sigma}\sb_{ijm}(\br-\br') \drho\sb_m(\br') d^3\br'   \right)
    \end{equation}
\end{widetext}
As a final comment regarding notation, we note the brackets represent an integration over $\br'$. Additionally, the quantity in brackets denotes all dependences on the line information. This notation will be utilized throughout the main text to refer to the Peach-Koehler functional.

\subsection{2. Line Integration of the cloud-in-cell weight function}
For reference, we give here the line integral of a straight line segment with respect to the cloud-in-cell weight function. In our previous work \cite{Anderson2021} we used a form of this that had a small error (fourth order in a parameter always less than 1), following a typographical error in \cite{Bertin2019}. It should be noted that the correct form is found in the preprint of the same \cite{Bertin2018arxiv}. However, we will restate the form of this line integral for clarity. 

We would like to integrate the dislocation configuration with respect to our cloud-in-cell weight function $w_L$ given by Eq. (\ref{eq:CIC}). In general, this evaluation will be centered at some $\br_D$; for brevity, we consider it to be centered at the origin. The support of the weight function $\Omega_L$ is given by a cube of side length $2L$ centered on the origin. This cube is composed of eight octants $\Omega_m$, cubes with side length $L$ with one corner on the origin.

Because the discrete dislocation configuration is composed of straight line segments ($\mathcal{L}=\bigcup_q \lambda_q$), we represent the dislocations which contribute to the line integral at this point by $\lambda_p^{(L)}$. Each segment $\lambda$ can be expressed in parametric form in terms of its origin $\boldsymbol{x}$, endpoint $\boldsymbol{x}+\boldsymbol{t}$, and a dimensionless parameter $a\in[0,1]$ as:
\begin{equation}
    \br^\lambda(a):=\boldsymbol{x}+a\boldsymbol{t}.
\end{equation}
It then remains to compute the integral of the weight function along this line.

In order to integrate the absolute value functions present, we use $\Omega_m$ to break the line into segments on which the sign of the three coordinates remains constant. Letting $a_m^\pm$ represent the line parameters at which $\lambda$ enters and leaves $\Omega_m$,\footnote{The line parameters of intersection with a rectangular grid can be efficiently computed using Siddon's algorithm \cite{Jacobs1998}. } we can represent by $s_i^m=\pm 1$ the sign of the coordinate $r_i(a)$ in the range $a\in[a_m^-,a_m^+]$. Utilizing the following expansion of the cloud-in-cell function [Eq. (\ref{eq:CIC})],
\begin{widetext}

\begin{equation}
    w_L(\boldsymbol{x},\boldsymbol{s}):=\frac{1}{L_1L_2L_3}\left[1-\sum_i \frac{s_i}{L_i} x_i + \sum_{i < j} \frac{s_i s_j}{L_iL_j} x_i x_j + \frac{s_1 s_2 s_3}{L_1L_2L_3}  x_1 x_2 x_3\right],
\end{equation}
the integral along this portion of the line can be evaluated as follows.
\begin{align}
    I_m &:= |\boldsymbol{t}| \int_{a_m^-}^{a_m^+} w_L(\br^\lambda(a),\boldsymbol{s}^m) da \\
    &= |\boldsymbol{t}|(a_m^+-a_m^-)-|\boldsymbol{t}|\sum_i \frac{s_i}{L_i} A_i + |\boldsymbol{t}|\sum_{i < j} \frac{s_i s_j}{L_iL_j} B_{ij} + \frac{s_1 s_2 s_3}{L_1L_2L_3} C_{123},
\end{align}
where we individually compute the various coordinate product integrals:
\begin{align}
    A_i &:=\int_{a_m^-}^{a_m^+}(x_i + at_i) da\\
        &=\left[ax_i + \frac{1}{2}a^2t_i \right]_{a_m^-}^{a_m^+}    \\
    B_{ij} &:=\int_{a_m^-}^{a_m^+}(x_i + at_i)(x_j + at_j) da\\
        &=\int_{a_m^-}^{a_m^+}(x_ix_j + a(x_it_j + x_j t_i)+a^2 t_i t_j) da\\
        &=\left[ax_i + \frac{1}{2}a^2(x_it_j + x_j t_i) + \frac{1}{3}a^3t_i t_j \right]_{a_m^-}^{a_m^+}\\
    C_{ijk} &:=\int_{a_m^-}^{a_m^+}(x_i + at_i)(x_j + at_j)(x_k + at_k) da\\
        &=\int_{a_m^-}^{a_m^+} \left(x_ix_jx_k + a(x_ix_kt_j + x_jx_kt_i + x_ix_jt_k) + a^2(x_it_it_j + x_it_jt_k + x_jt_it_k) + a^3 t_it_jt_k\right)da\\
        &= \left[x_ix_jx_k + \frac{1}{2}a^2(x_ix_kt_j + x_jx_kt_i + x_ix_jt_k) + \frac{1}{3}a^3(x_it_it_j + x_it_jt_k + x_jt_it_k) + \frac{1}{4}a^4 t_it_jt_k\right]_{a_m^-}^{a_m^+}
\end{align}

\end{widetext}
Finally, this allows us to express the total line integral, given by
\begin{align}
    I_\lambda &= |\boldsymbol{t}| \int_0^1 w_L\left(\br^\lambda(a)\right)da \\
    &= \sum_{m=1}^8 I_m.
\end{align}

The total of the line integral for all segments passing through $\Omega_L$ gives a scalar line density field, while incorporating the line direction gives the vector density field:
\begin{align}
    \rho(\br_D;L) &=\sum_q I_{\lambda^{(L)}_q}, \\
    \brho(\br_D;L) &= \sum_q I_{\lambda^{(L)}_q}\frac{\boldsymbol{t}_{\lambda^{(L)}_q}}{|\boldsymbol{t}_{\lambda^{(L)}_q|}}.
\end{align}
If only a single segment passes through the volume, the scalar density and the norm of the vector density will be equal. In general, however, this is not the case, and the polarization $|\brho|/\rho$ is generally less than unity due to the presence of two or more straight line segments.

\section*{Acknowledgment}
This work was supported by the US Department of Energy, Office of Science, Division of Materials Sciences and Engineering, through award number DE-SC0017718 at Purdue University.

\bibliography{Corr_2_bib}
\end{document}